\documentclass[12pt,preprint]{aastex}
\usepackage{amsmath,amstext,amsbsy,amsopn,psfig,natbib,epsfig,sidecap}

\received{-}
\revised{-}

\slugcomment{}

\shorttitle{SDSS-II Supernovae Survey Spectroscopy}
\shortauthors{Zheng et al.}

\begin{document}

\title{First-Year Spectroscopy for the SDSS-II Supernova Survey}
\author{
Chen~Zheng,\altaffilmark{1}
Roger~W.~Romani,\altaffilmark{1}
Masao~Sako,\altaffilmark{2}
John~Marriner,\altaffilmark{3}
Bruce~Bassett,\altaffilmark{4,5}
Andrew~Becker,\altaffilmark{6}
Changsu Choi,\altaffilmark{7}
David~Cinabro,\altaffilmark{8}
Fritz~DeJongh,\altaffilmark{3}
Darren~L.~Depoy,\altaffilmark{9}
Ben~Dilday,\altaffilmark{11,10}
Mamoru~Doi,\altaffilmark{12}
Joshua~A.~Frieman,\altaffilmark{3,11,13}
Peter~M.~Garnavich,\altaffilmark{14}
Craig~J.~Hogan,\altaffilmark{6}
Jon~Holtzman,\altaffilmark{15}
Myungshin~Im,\altaffilmark{7}
Saurabh~Jha,\altaffilmark{1}
Richard~Kessler,\altaffilmark{11,16}
Kohki~Konishi,\altaffilmark{17}
Hubert~Lampeitl,\altaffilmark{18}
Jennifer~L.~Marshall,\altaffilmark{9}
David~McGinnis,\altaffilmark{3}
Gajus~Miknaitis,\altaffilmark{3}
Robert~C.~Nichol,\altaffilmark{19}
Jose~Luis~Prieto,\altaffilmark{9}
Adam~G.~Riess,\altaffilmark{18,20}
Michael~W.~Richmond,\altaffilmark{21}
Donald~P.~Schneider,\altaffilmark{22}
Mathew~Smith,\altaffilmark{19}
Naohiro~Takanashi,\altaffilmark{12}
Kouichi~Tokita,\altaffilmark{12}
Kurt~van~der~Heyden,\altaffilmark{5}
Naoki~Yasuda,\altaffilmark{17}
Roberto~J.~Assef,\altaffilmark{9}
John~Barentine,\altaffilmark{23,24}
Ralf~Bender,\altaffilmark{25,26}
Roger~D.~Blandford,\altaffilmark{1}
Malcolm~Bremer,\altaffilmark{27}
Howard~Brewington,\altaffilmark{24}
Chris~A.~Collins,\altaffilmark{28}
Arlin~Crotts,\altaffilmark{29}
Jack~Dembicky,\altaffilmark{24}
Jason~Eastman,\altaffilmark{9}
Alastair~Edge,\altaffilmark{30}
Ed~Elson,\altaffilmark{4,5}
Michael~E.~Eyler,\altaffilmark{31}
Alexei~V.~Filippenko,\altaffilmark{32}
Ryan~J.~Foley,\altaffilmark{32}
Stephan~Frank,\altaffilmark{9}
Ariel Goobar,\altaffilmark{33}
Michael~Harvanek,\altaffilmark{24,34}
Ulrich~Hopp,\altaffilmark{25,26}
Yutaka~Ihara,\altaffilmark{12}
Steven~Kahn,\altaffilmark{1}
William~Ketzeback,\altaffilmark{24}
Scott~J.~Kleinman,\altaffilmark{24,35}
Wolfram~Kollatschny,\altaffilmark{36}
Jurek~Krzesi\'{n}ski,\altaffilmark{24,37}
Giorgos Leloudas,\altaffilmark{38}
Daniel~C.~Long,\altaffilmark{24}
John~Lucey,\altaffilmark{30}
Elena~Malanushenko,\altaffilmark{24}
Viktor~Malanushenko,\altaffilmark{24}
Russet~J.~McMillan,\altaffilmark{24}
Christopher~W.~Morgan,\altaffilmark{9,31}
Tomoki~Morokuma,\altaffilmark{12,39}
Atsuko~Nitta,\altaffilmark{24,40}
Linda Ostman,\altaffilmark{33}
Kaike~Pan,\altaffilmark{24}
A.~Kathy~Romer,\altaffilmark{41}
Gabrelle~Saurage,\altaffilmark{24}
Katie~Schlesinger,\altaffilmark{9}
Stephanie~A.~Snedden,\altaffilmark{24}
Jesper~Sollerman,\altaffilmark{38,42}
Maximilian~Stritzinger,\altaffilmark{38}
Linda~C.~Watson,\altaffilmark{9}
Shannon~Watters,\altaffilmark{24}
J.~Craig~Wheeler,\altaffilmark{23}
and
Donald~York\altaffilmark{13,16}
}

\altaffiltext{1}{
Kavli Institute for Particle Astrophysics and Cosmology, 
Stanford University, Stanford, CA 94305-4060.
}
\altaffiltext{2}{
Department of Physics and Astronomy,
University of Pennsylvania, 203 South 33rd Street, Philadelphia, PA 19104.
}
\altaffiltext{3}{
Center for Particle Astrophysics, 
Fermi National Accelerator Laboratory, P.O. Box 500, Batavia, IL 60510.
}
\altaffiltext{4}{
Department of Mathematics and Applied Mathematics, 
University of Cape Town, Rondebosch 7701, South Africa.
}
\altaffiltext{5}{
South African Astronomical Observatory, 
P.O. Box 9, Observatory 7935, South Africa.
}
\altaffiltext{6}{
Department of Astronomy, 
University of Washington, Box 351580, Seattle, WA 98195.
}
\altaffiltext{7}{
Department of Physics \& Astronomy, FPRD, 
Seoul National University, Seoul, South Korea.
}
\altaffiltext{8}{
Department of Physics, 
Wayne State University, Detroit, MI 48202.
}
\altaffiltext{9}{
Department of Astronomy, 
Ohio State University, 140 West 18th Avenue, Columbus, OH 43210-1173.
}
\altaffiltext{10}{
Department of Physics, 
University of Chicago, Chicago, IL 60637.
}
\altaffiltext{11}{
Kavli Institute for Cosmological Physics, 
The University of Chicago, 5640 South Ellis Avenue Chicago, IL 60637.
}
\altaffiltext{12}{
Institute of Astronomy, Graduate School of Science, 
University of Tokyo 2-21-1, Osawa, Mitaka, Tokyo 181-0015, Japan.
}
\altaffiltext{13}{
Department of Astronomy and Astrophysics, 
The University of Chicago, 5640 South Ellis Avenue, Chicago, IL 60637.
}
\altaffiltext{14}{
University of Notre Dame, 225 Nieuwland Science, Notre Dame, IN 46556-5670.
}
\altaffiltext{15}{
Department of Astronomy, 
MSC 4500, 
New Mexico State University, P.O. Box 30001, Las Cruces, NM 88003.
}
\altaffiltext{16}{
Enrico Fermi Institute, 
University of Chicago, 5640 South Ellis Avenue, Chicago, IL 60637.
}
\altaffiltext{17}{
Institute for Cosmic Ray Research, 
University of Tokyo, 5-1-5, Kashiwanoha, Kashiwa, Chiba, 277-8582, Japan.
}
\altaffiltext{18}{
Space Telescope Science Institute, 
3700 San Martin Drive, Baltimore, MD 21218.
}
\altaffiltext{19}{
Institute of Cosmology and Gravitation, 
Mercantile House, 
Hampshire Terrace, University of Portsmouth, Portsmouth PO1 2EG, UK.
}
\altaffiltext{20}{
Department of Physics and Astronomy, 
Johns Hopkins University, 3400 North Charles Street, Baltimore, MD 21218.
}
\altaffiltext{21}{
Physics Department, 
Rochester Institute of Technology, 
85 Lomb Memorial Drive, Rochester, NY 14623-5603.
}
\altaffiltext{22}{
Department of Astronomy and Astrophysics, 
The Pennsylvania State University, 
525 Davey Laboratory, University Park, PA 16802.
}
\altaffiltext{23}{
Department of Astronomy, 
McDonald Observatory, University of Texas, Austin, TX 78712.
}
\altaffiltext{24}{
Apache Point Observatory, P.O. Box 59, Sunspot, NM 88349.
}
\altaffiltext{25}{
Universitaets-Sternwarte Munich, 1 Scheinerstr, Munich, D-81679, Germany.
}
\altaffiltext{26}{
Max Planck Institute for Extraterrestrial Physics, D-85748, 
Garching, Munich, Germany.
}
\altaffiltext{27}{
H. H. Wills Physics Laboratory, University of Bristol, Bristol, BS8 1TL, UK.
}
\altaffiltext{28}{
Astrophysics Research Institute, Liverpool John Moores University, 
Birkenhead CH41 1LD, UK.
}
\altaffiltext{29}{
Department of Astronomy, 
Columbia University, New York, NY 10027.
}
\altaffiltext{30}{
Department of Physics, University of Durham, South Road, 
Durham, DH1 3LE, UK.
}
\altaffiltext{31}{
Department of Physics, United States Naval Academy, 
572C Holloway Road, Annapolis, MD 21402.
}
\altaffiltext{32}{
Department of Astronomy, University of California, Berkeley, CA 94720-3411.
}
\altaffiltext{33}{
Physics Department, Stockholm University, AlbaNova University Center, 
106 91 Stockholm, Sweden.
}
\altaffiltext{34}{
Lowell Observatory, 1400 Mars Hill Rd., Flagstaff, AZ 86001.
}
\altaffiltext{35}{
Subaru Telescope, 650 N. A'Ohoku Place, Hilo, HI 96720.
}
\altaffiltext{36}{
Institut f\"{u}r Astrophysik, Universit\"{a}t G\"{o}ttingen, 
Friedrich-Hund-Platz 1, D-37077 G\"{o}ttingen, Germany
}
\altaffiltext{37}{
Obserwatorium Astronomiczne na Suhorze, 
Akademia Pedagogicazna w Krakowie, 
ulica Podchor\c{a}\.{z}ych 2, PL-30-084 Krak\'{o}w, Poland.
}
\altaffiltext{38}{
Dark Cosmology Centre, Niels Bohr Institute, University of Copenhagen, 
DK-2100, Denmark.
}
\altaffiltext{39}{
National Astronomical Observatory of Japan, 2-21-1, Osawa, Mitaka, 
Tokyo 181-8588, Japan.
}
\altaffiltext{40}{
Gemini Observatory, 670 North A'ohoku Place, Hilo, HI 96720.
}
\altaffiltext{41}{
Astronomy Center, University of Sussex, Falmer, Brighton BN1 9QJ, UK.
}
\altaffiltext{42}{
Astronomy Department, Stockholm University, AlbaNova University Center, 
106 91 Stockholm, Sweden.
}

\begin{abstract}

This paper presents spectroscopy of supernovae discovered in the first season
of the Sloan Digital Sky Survey-II Supernova Survey. This program searches for
and measures multi-band light curves of supernovae in the redshift range $z =
0.05 - 0.4$, complementing existing surveys at lower and higher redshifts. Our
goal is to better characterize the supernova population, with a particular
focus on SNe Ia, improving their utility as cosmological distance indicators
and as probes of dark energy. Our supernova spectroscopy program features
rapid-response observations using telescopes of a range of apertures, and
provides confirmation of the supernova and host-galaxy types as well as
precise redshifts.  We describe here the target identification and
prioritization, data reduction, redshift measurement, and classification of
$129$ SNe Ia, $16$ spectroscopically probable SNe Ia, $7$ SNe Ib/c, and $11$
SNe II from the first season.  We also describe our efforts to measure and
remove the substantial host galaxy contamination existing in the majority of
our SN spectra.

\end{abstract}


\maketitle

\section{INTRODUCTION}

During the last few decades, studies of Type Ia supernovae (SNe Ia) have made
significant contributions to our understanding of cosmology. In particular, SN
Ia samples from low redshifts ($z \lesssim 0.1$) have helped to constrain the
Hubble constant \citep{ha96, jh99, jh07}. Their comparison with SN Ia samples
from high redshifts ($z \gtrsim 0.4 $) has led to evidence for an accelerating
universe \citep[][see \citet{fi05b} for a review]{rie98, per99}.  The key is
that these SNe prove to be remarkably homogeneous ``standard candles'' after
correcting for the empirical luminosity vs. decline-rate relation
\citep[e.g.,][]{ph93}. However, surveys covering large areas with intermediate
depth have proved difficult to perform, leading to a ``redshift desert"
\citep[$0.1 \lesssim z \lesssim 0.4$;][]{ri04} in the SN Ia Hubble
diagram. Further, merging data from multiple surveys raises questions of
systematics control; high-precision cross calibration is required to
differentiate between cosmological models. To further calibrate the supernova
luminosities, to probe for possible redshift evolution, and to study the
empirical properties of the ``dark energy'' that has been invoked to account
for the acceleration of the universe, it is critical to fill in this redshift
desert with well-studied data and to connect the low-$z$ and high-$z$
populations.

The Supernova Survey of the Sloan Digital Sky Survey-II (SDSS-II), comprising
three three-month campaigns (September through November of 2005 -- 2007), was
launched to discover SNe Ia and acquire photometric and spectroscopic
observations in this sparsely sampled intermediate redshift interval of $z =
0.05$--0.4 \citep{fr07}. The large (300 square degrees), moderately deep ($r
\approx 22.5$~mag) survey provides a significant volume for untargeted
discovery of these intermediate-redshift supernovae, thus complementing
ongoing low-$z$ surveys and follow-up programs (e.g., Lick Observatory
Supernova Search -- LOSS\footnotemark, \citealt{li00, fi01, fi05a}; Carnegie
Supernova Program -- CSP\footnotemark, \citealt{ha06}; Nearby Supernova
Factory -- SNFactory\footnotemark, \citealt{co06}; the Center for Astrophysics
follow-up effort -- CfA SN Group\footnotemark,
\citealt{rie95,rie96,rie99,jh06,jh07}).  The survey is sufficiently deep to
overlap with the high-$z$ samples (e.g., Canada-France-Hawaii Telescope
Supernova Legacy Survey - SNLS\footnotemark, \citealt{as06}; Equation of
State: SupErNovae trace Cosmic Expansion -- ESSENCE\footnotemark,
\citealt{mi07, wo07}) and help to explore the kinematics of the expanding
universe during the interval when most models of dark energy anticipate
maximum departure from a simple, flat, $\Lambda$-Cold Dark Matter (LCDM)
cosmology. The uniformity of the SDSS photometric calibration system provides
precise measurements of supernova light curves in five filters
\citep[$ugriz$;][]{fu96}. The large survey volume also facilitates discovery
of rare supernova types, sampling the full extent of the SN Ia population, as
well as allowing searches for SNe Ib/c, SNe II, hypernovae and other peculiar
denizens of the astronomical zoo.

\footnotetext[1]{ http://astro.berkeley.edu/$\sim$bait/kait.html .}
\footnotetext[2]{ http://csp1.lco.cl/$\sim$cspuser1/PUB/CSP.html .}
\footnotetext[3]{ http://snfactory.lbl.gov/ .}
\footnotetext[4]{ http://www.cfa.harvard.edu/cfa/oir/Research/supernova/SNgroup.html .}
\footnotetext[5]{ http://www.cfht.hawaii.edu/SNLS .}
\footnotetext[6]{ http://www.ctio.noao.edu/essence .}

This paper describes our spectroscopic follow-up techniques and presents
results from spectroscopic observations of supernovae discovered in the first
season of the SDSS-II Supernova Survey. During the Fall 2005 campaign, we
repeatedly scanned SDSS stripe $82$, with alternate observations of the
northern and southern strips, maintaining full coverage of the
2.5$^\circ$-wide stripe on a cadence of $\sim 2$~d. Supernova candidates were
identified by rapid on-mountain processing in the $g$, $r$, and $i$
filters. Targets that passed a variety of quality control cuts were then
inspected by humans and prioritized for spectroscopic follow-up observations
\citep{sa07}.  The primary spectroscopic facilities used to complement the
imaging survey were the $2.4$~m Hiltner telescope at MDM, the $3.5$~m
Astronomy Research Consortium (ARC) telescope at Apache Point Observatory, and
the queue-scheduled $9.2$~m Hobby-Eberly Telescope (HET) at McDonald
Observatory. Individual observing campaigns were also conducted with the
$4.2$~m William Herschel Telescope (WHT), the $8.2$m Subaru, and the $10$~m
Keck~I telescope. A total of 259 spectra were obtained during the first
season, yielding $129$ spectroscopically confirmed SNe Ia with a wide range of
epochs ($-12$\,d to $54$\,d relative to $B$-band maximum light), 16
spectroscopically probable SNe Ia, and a handful of other supernova types,
including peculiar SNe Ia and broad-lined SN Ib/c ``hypernovae.'' Several
objects were observed at multiple epochs.

In this paper, we present our spectroscopic data reduction and analysis
methods. We briefly summarize the candidate identification and prioritization
algorithms (\S2), and we describe the spectroscopic observations at several
follow-up telescopes as well as the basic data reduction (\S3). In \S4, we
describe the procedures used for supernova classification and redshift
determination, including efforts to measure and remove spectral contamination
by the host-galaxy light. Section~5 summarizes the basic results, describing a
few of the peculiar spectra. A complete analysis of the spectral features in
the SN~Ia population and extension of the analysis to Seasons 2 and 3 are
deferred to later publications. However, in \S6, we note potential results
from the analysis of the full spectroscopy data and summarize our present
conclusions.

\section{SELECTION OF SUPERNOVA CANDIDATES FOR SPECTROSCOPY}

The SDSS-II Supernova Program uses a CCD camera \citep{gu98} mounted on the
Apache Point Observatory (APO) 2.5-m telescope \citep{gu06} to obtain repeated
scans of a 300 square degree region aligned along the celestial equator in the
Southern Galactic Hemisphere. Images in this \hbox{$2.5^{\circ} \times
120^{\circ}$} area (SDSS Stripe 82) are obtained almost simultaneously using
five filters \citep[$ugriz$;][]{fu96}. The images have sufficient sensitivity
to discover SNe Ia out to $z \approx 0.4$, while providing high-quality light
curves \citep{ho07} for a large fraction of the supernovae on an accurately
calibrated photometric system \citep[$\sim$1\%; ][]{iv07}.  Technical
summaries of the SDSS and the data products can be found in \cite{yo00},
\cite{ad07}, and references therein. An overview of the SDSS-II supernova
program is presented by \cite{fr07}.

\begin{figure}
\begin{center}
\includegraphics[angle=0,scale=.5]{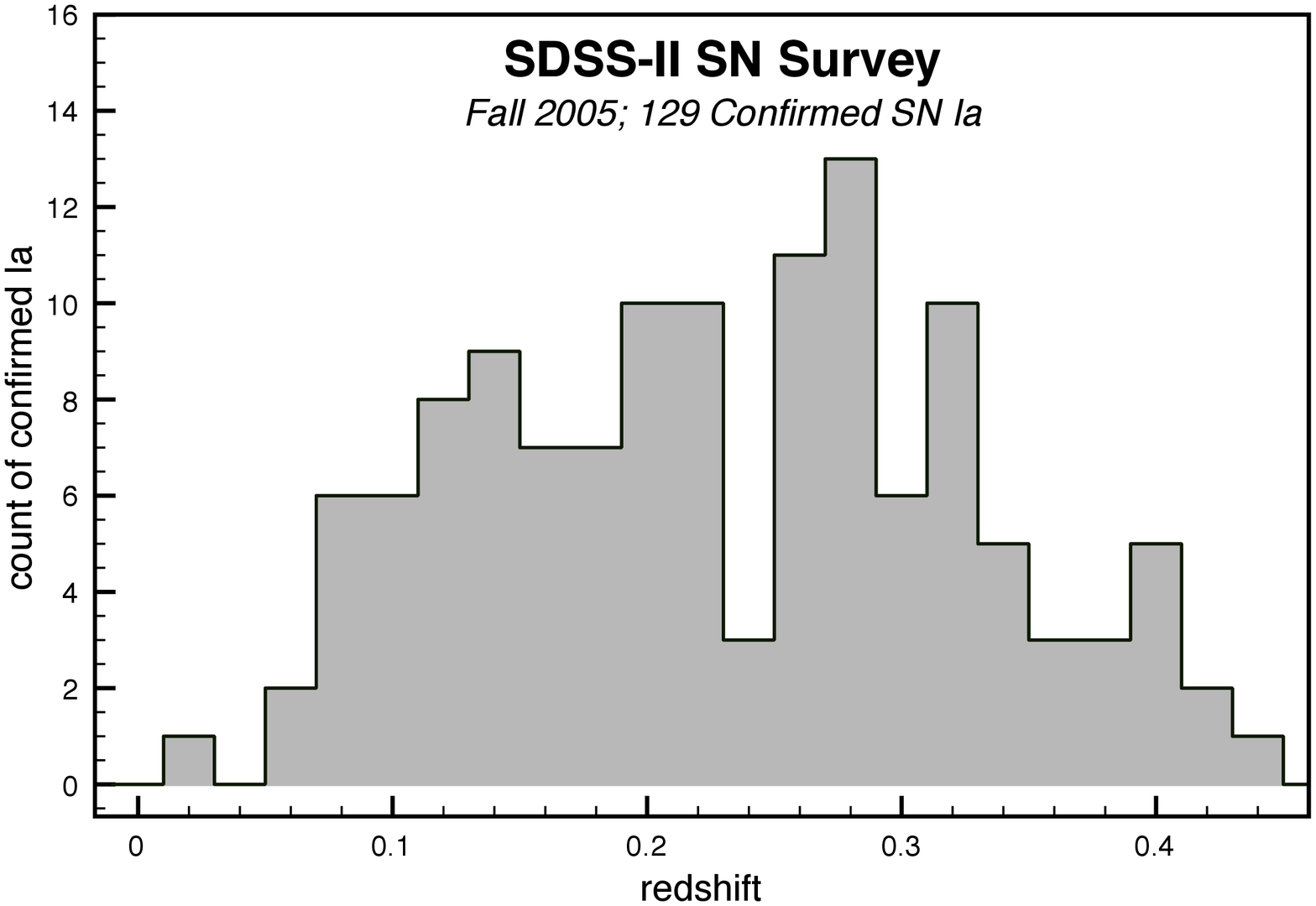}
\caption{The redshift distribution of $129$ spectroscopically confirmed SNe Ia
  in 2005. \label{fig:count}}
\includegraphics[angle=0,scale=.5]{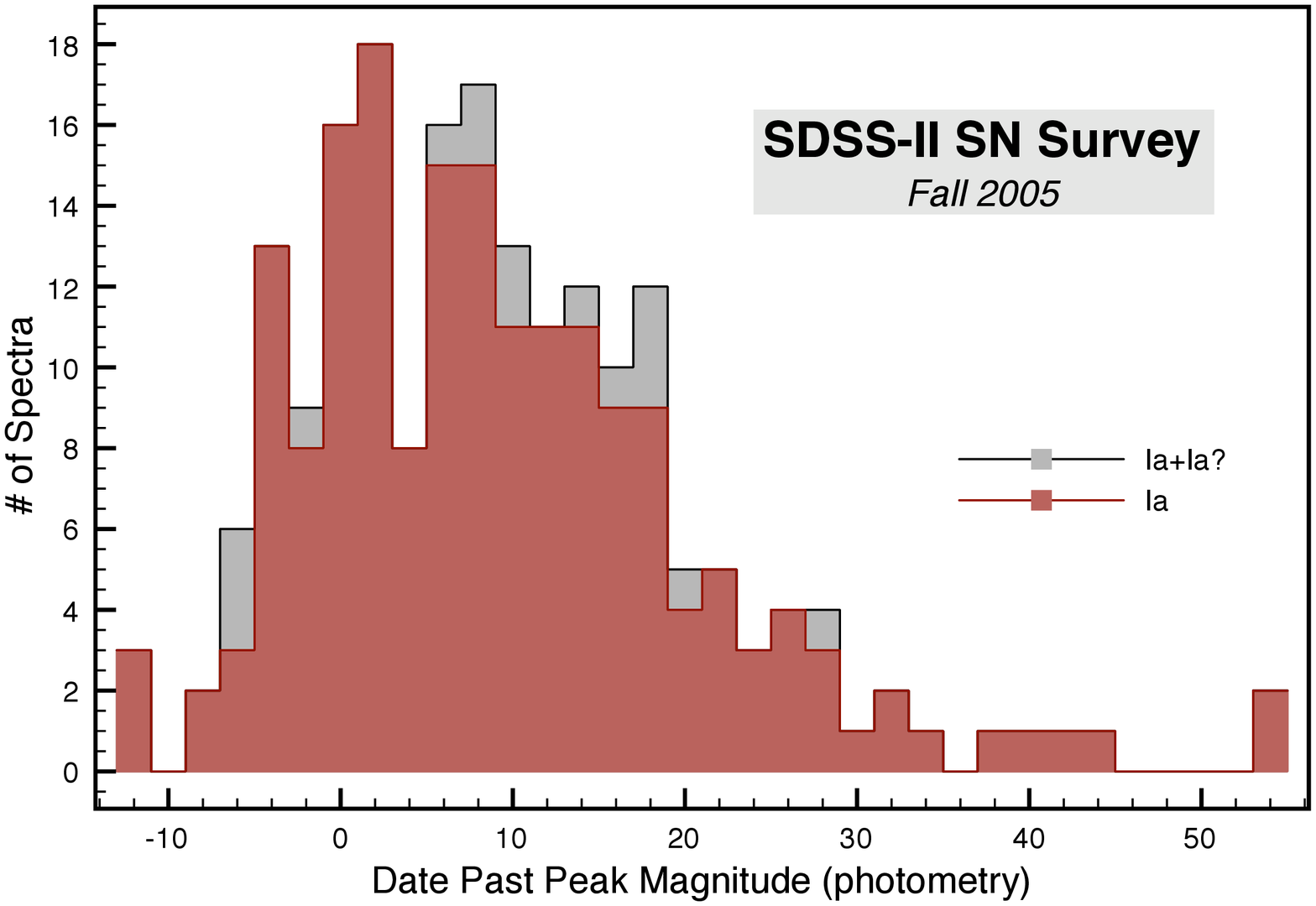}
\caption{Distribution of first spectroscopy epoch, relative to $B$-band
  maximum light, for spectroscopically confirmed and probable SNe Ia; most
  spectra were obtained before $+20$~d. \label{fig:epoch}}
\end{center}
\end{figure}

With the large data volume produced for each clear night, it was necessary to
flag supernova candidates using a dedicated computer cluster at Apache Point
Observatory. Co-added template images from previous years were registered and
convolved to match the search images in the $g$, $r$, and $i$
bands. Statistically significant variable sources appearing in two or more
filters were flagged. Within 24~hours, thousands of new detections from each
night of observation were transferred to Fermilab and entered into the
supernova database. These detections (or ``objects'') were visually inspected
to identify possible supernova candidates. Host-galaxy properties (morphology,
color, and photometric redshift) were assembled along with estimates for the
supernova photometric redshift, extinction, date of maximum light, photometric
type, and stretch factor, as obtained from the preliminary photometry.  For
all active candidates, these parameters were updated daily. Visual grading of
the candidates produced a cleaner supernova set. Efficiencies for selecting
true supernovae were actively monitored with a large number of ``fake''
supernovae added to the data stream. Details of the candidate identification
and characterization are described by \cite{sa07}.

With such a large number of candidate supernovae and limited spectroscopic
resources, only the ``best'' objects were subject to follow-up spectroscopic
observations. In particular, during the first season, we focused (with a few
exceptions) on objects for which a classification of SN~Ia was strongly
preferred by the initial photometry. We also selected for large SN-galaxy
separation, minimal photometric evidence for dust extinction, and interesting
SNe (those with lowest or highest redshift, underluminous hosts, and peculiar
light curves). This approach certainly biases the distribution of supernova
properties, except for $z \lesssim 0.12$, where we attempted to make the data
as complete as possible \citep{di07}. These ``best'' supernovae were assigned
to the various telescopes: APO, MDM, and WHT for the brighter objects with
lower estimated redshifts, or HET and Subaru for the fainter distant
objects. Significant efforts were made to avoid duplicate observations and to
maximize the observation efficiency of each telescope. Only a modest fraction
($\sim 25\%$) of the selected candidates were not observed within $20$
rest-fame days of SN maximum or yielded noisy spectra; host-galaxy spectra and
redshifts are being measured for a sample of those objects missed due to poor
weather or oversubscribed spectroscopic resources.

\section{OBSERVATIONS AND REDUCTIONS}

During Fall 2005, three telescopes were routinely available for SDSS-II
supernova spectroscopy: MDM $2.4$~m ($49$ shared nights), ARC $3.5$~m ($31$
half-nights), and the HET $9.2$~m ($64.5$ hours of queue time). Several
additional dedicated campaigns helped greatly, with $6$ nights on the $4.2$~m
WHT, $6$ shared nights on Subaru, and one target-of-opportunity night on Keck.
A handful of additional spectra were obtained at other facilities based on the
SDSS-II supernova candidate announcements. In total, $259$ useful spectra were
obtained which yielded $129$ spectroscopically confirmed SNe Ia, $16$ probable
SNe Ia, $7$ SNe Ib/c, and $11$ SNe II. Figure \ref{fig:count} shows the
redshift distribution of the confirmed SNe Ia. The distribution is
approximately flat for $0.1< z < 0.35$ with a deficit at $z \approx 0.2$. This
deficit is almost purely a result of bias in our follow-up selection, with
bright low-$z$ sources assigned to the smaller telescopes and distant sources
targeted at the $\ge 8$~m facilities.

Most supernovae had spectroscopy within 20~d of photometric maximum (Figure
\ref{fig:epoch}). Multi-epoch follow-up spectroscopy was carried out for the
peculiar Type Ia \object{SN 2005js} \citep[$2$ epochs; similar to SN 1991bg --
  e.g.,][]{fi92b}, the underluminous, low expansion velocity Type Ia
\object{SN 2005hk} \citep[$10$ epochs; see also][]{ch06, ph07}, and a peculiar
event, \object{SN 2005gj} \citep[$23$ epochs; see also][]{al06, pr07}, which
had broad SN~Ia-like features with superimposed hydrogen emission lines
\citep[similar to \object{SN 2002ic},][]{ha03}.  Three broad-lined Type Ic
``hypernovae'' \citep[similar to \object{SN 1998bw} and \object{SN 2002ap};
  e.g.,][]{sta00,fol03} were observed as well: \object{SN 2005fk}, \object{SN
  2005kr}, and \object{SN 2005ks}, selected as Type Ic ``hypernovae'' based on
their light curves, each had a \object{SN 2002ap}-like spectrum. At least some
``hypernovae'' are believed to be associated with gamma-ray bursts
\citep[e.g.,][]{st03,ma03}.

Here we summarize the observing configurations for the main facilities. The
individual targeted supernova positions and dates of observations are listed
in Table~\ref{tbl:obs}.

\begin{figure}
\begin{center}
\includegraphics[angle=0,scale=.60]{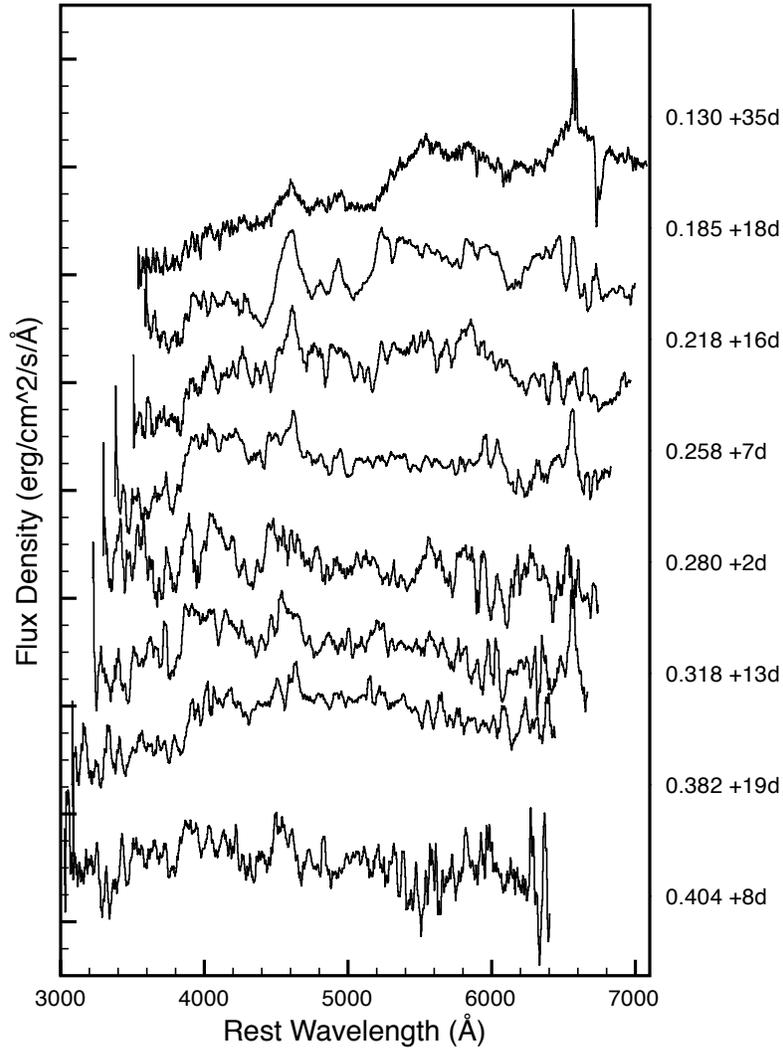}
\caption{A sample of observed Ia spectra from the Fall 2005 run with significant host-galaxy contamination. All spectra have been shifted to the rest frame (numbers at right give the redshifts and epochs relative to $B$-band maximum) and have been normalizedfor display purposes. The top spectrum (from Subaru) has not been corrected for telluric absorption.
\label{fig:hostcontam}}
\end{center}
\end{figure}

\subsection{Telescopes}

\subsubsection{MDM 2.4~m}
The MDM $2.4$~m telescope was used for both imaging and spectroscopy of the SN
candidates. With a large number of nights and an efficient spectrograph, this
facility made important contributions despite its modest aperture. Spectra
were taken with the Boller \& Chivens CCD Spectrograph (CCDS), with a $150$
l~mm$^{-1}$ grating ($4700$~\AA\ blaze) feeding a $1200\times800$ Loral CCD. With a
$2''$ slit we obtained a resolution of $15$~\AA\ covering $\sim$3800--7300~\AA. 
Observations with the fixed North/South slit were taken within 1~hr of
transit, whenever possible, to minimize refractive slit losses \citep{fi82}. Typically the
exposure was $3\times900$~s. A total of $38$ spectra were obtained confirming
$16$ SNe Ia, $2$ SNe Ib/c, and $4$ host galaxies, with a redshift distribution
centered at $z \approx 0.08$. This telescope was used extensively to perform
multi-epoch spectroscopy of bright, nearby events.

\subsubsection{ARC 3.5~m}
The $54$ ARC spectra were obtained with the Dual Imaging Spectrograph (DIS),
which has dual (red/blue) cameras with Marconi $2048\times1024$ pixel
back-illuminated chips. The blue spectra used a $300$ l~mm$^{-1}$ grating giving
2.43~\AA\ pixel$^{-1}$ centered at $4224$~\AA; the red camera employed a $300$
l~mm$^{-1}$ grating for 2.26~\AA\ pixel$^{-1}$ dispersion, centered at $7500$~\AA. 
With a slit width of $\sim 1.5^{\prime\prime}$, the effective resolution was 
$\sim$8--9~\AA. For most observations with a visible host, the slit was aligned
with the supernova and galaxy core. Typical exposures were 300--900~s and
3--5 exposures were combined for the final spectra. The observations yielded
$38$ SNe ($29$ SNe Ia, $3$ probable SNe Ia, $5$ SNe II, and $1$ SN Ib), with a
median redshift of $\sim$0.11.

\subsubsection{HET 9.2~m}
The HET was used to observe supernova candidates with higher photo-$z$.
Exposures were made with the Marcario Low Resolution Spectrograph
\citep[LRS;][]{hi98}, employing a Ford Aerospace $3072\times1024$ pixel chip and the
$300$ l~mm$^{-1}$ G1 grism for a dispersion of $5$~\AA\ pixel$^{-1}$. With a slit width of
$2''$, the effective resolution was $\sim 20$~\AA. Observations were made
with the slit at the parallactic angle \citep{fi82} to avoid differential slit losses and
improve (relative) spectrophotometry across the range 4067--10700~\AA. As
the HET is queue scheduled \citep{sh07}, many supernova spectra were obtained
soon after discovery, and lower-priority observation time was spent 
acquiring late-time data for supernovae with previous spectroscopic confirmation.
We obtained $92$ HET spectra over a period of $69$ days, yielding $61$ SNe
($45$ SNe Ia, $10$ probable SNe Ia, $1$ SN Ib, $2$ SNe II, and $3$ SNe Ibc)
with a redshift distribution centered at $z \approx 0.27$. A few multi-epoch
observations and a few unclassifiable spectra completed the sample.

\subsubsection{WHT 4.2~m}
Six nights of observing were obtained with the ISIS Double Beam Spectrograph.
Spectra with a combined red/blue coverage of 3900--8900~\AA\ were
obtained for $30$ supernova candidates. Typically the slit width was set at
$1''$, providing a resolution of $4.3$~\AA\ in the blue and $7.5$~\AA\ in the
red, although the slit was adjusted to accommodate variable seeing. For the
observations in October and November, a GG$495$ filter was employed in the red
channel to eliminate second-order contamination. The observations provided
$26$ SNe Ia, $3$ probable SNe Ia, and $1$ SN Ic (with a median supernova
redshift of $\sim 0.16$).

\subsubsection{Subaru 8.2~m}
At this telescope we used the Faint Object Camera and Spectrograph (FOCAS),
taking advantage of the excellent image quality to employ a $0.8''$ slit and
target higher-redshift SNe with significant host-galaxy contamination. Separate
exposures were made using blue and red $300$ l~mm$^{-1}$ grisms. The blue setting
covered 3650--6000~\AA, while the red covered 4900--9000~\AA. The
resulting resolution was $\sim$8--12~\AA. Typical exposures of $3\times
300$~s allowed us to confirm $30$ SNe Ia, 1 probable SN Ia and $2$ SNe II, with
a median redshift of $\sim$0.25.

\subsubsection{Keck 10-m}
One target-of-opportunity night (2005 November 5) was obtained with the Low
Resolution Imaging Spectrograph (LRIS; Oke et al. 1995) on Keck I. The blue side covered 
$\sim$3200--5700~\AA\ with a $600$ l~mm$^{-1}$ grating ($4000$~\AA\ blaze; 1.92~\AA\
pixel$^{-1}$ dispersion), and the red side covered $\sim$5500--9400~\AA\ with a $400$
l~mm$^{-1}$ grating ($8500$~\AA\ blaze; 0.61~\AA\ pixel$^{-1}$ dispersion). A slit of
width $1''$ was oriented at the parallactic angle for an effective resolution of
$8.9$~\AA\ (blue) and $4.5$~\AA\ (red). With exposures ranging from 700 to
2400~s, we observed $12$ SNe Ia and one SN IIn; the
median redshift was $\sim$0.24. Most these objects had been previously
confirmed as spectroscopic SNe~Ia at our other telescopes, but the Keck spectra
generally had higher signal-to-noise ratio (S/N) and provided important
rest-frame UV information on the supernovae.

\subsection{Calibrations and Reductions}

Basic data calibration (bias/overscan subtraction, flux calibration, and
wavelength calibration) was performed by the individual observing teams using
standard IRAF\footnote{IRAF is distributed by the National Optical Astronomy 
Observatories, which are operated by the Association of Universities for 
Research in Astronomy, Inc., under cooperative agreement with the National
Science Foundation.} routines. Supernova spectra were extracted from the
geometrically corrected two-dimensional (2D) spectra, often with significant amounts of 
host-galaxy background (see Figure \ref{fig:hostcontam}); for higher-$z$ targets embedded in
their hosts, modest spatial-width apertures were used in extracting the supernova
spectra to minimize host contamination. In half of the cases, a separate
host-dominated aperture could be extracted from the 2D spectrum to produce 
a host-galaxy spectrum. For ARC observations, a separate host-core aperture was often extracted. For the other observations, the near parallactic orientation minimized chromatic slit
losses; however, given our narrow slits and extraction apertures, no attempt at
absolute spectrophotometry was made. Reductions of the HET, ARC, WHT, and Keck data
included a correction for telluric absorption.

\section{SN IDENTIFICATION}
\begin{figure}
 \centering
 \includegraphics[angle=0,scale=.60]{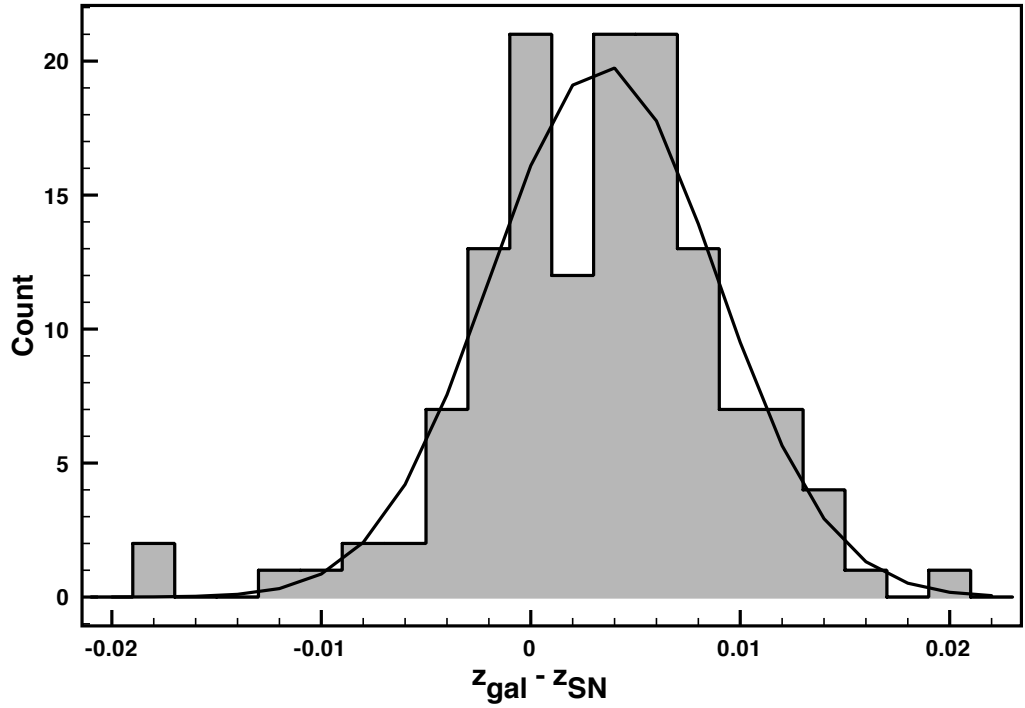}
 \caption{Histogram of $z_g-z_{SN}$ for 136 spectra of 116 SNe in our 2005 data 
with both host-galaxy and SN redshift solutions. The red line shows a 
best-fit Gaussian, with offset $\delta z = 0.003$ and a width of 
$\sigma_z = 0.005$. The latter is used as an estimate of the error in the 
SN redshift when narrow host-galaxy lines are not detected.
 \label{fig:zgzsn}}
\end{figure}

The principal goal of the spectroscopic observations is to classify the SN and
measure its redshift.  We can also use these measurements to improve the
choice of future spectroscopic targets. For low-$z$ supernovae with high S/N
spectra, the classification is, in principle, simple \citep{fi97} -- Type II
show strong Balmer lines, Type Ia show strong Si II lines, Type Ib lack H
lines but show strong He emission, Type Ic lack all of the above. Traditional
classification is done by visual inspection. In practice, however, our spectra
have widely varying S/N, cover a large redshift range, and suffer significant
host-galaxy contamination and extinction. Further, with this large, relatively
unbiased sample, we expect rare, ``peculiar'' SNe to be present (which are not
discussed further in the present paper, but will be part of a future analysis
with the complete spectroscopic sample). We have sought to develop automatic
classification programs capable of dealing with our sample, exploring template
cross-correlation and principal component analysis (PCA) techniques. While
these have substantial discriminating power and also serve to find best-fit
parameters and their associated statistical errors, the final object
(supernova- and host-type) classification required some human judgment in at
least half of the cases.

Of course, we also need to obtain the accurate redshifts required when using
the supernovae for cosmology and population studies. These were
robustly measured in the cross-correlation analysis, especially in the $\sim
50$\% of the spectra with large host-galaxy contamination or separate host spectra.
These are augmented by ``host-only'' spectra, obtained for supernovae that had
faded before spectroscopy could be arranged (these sources are not discussed
further in the present paper). When cross correlation could be locked to the
narrow-line host features, we refer to the result as a ``galaxy redshift'';
these have a typical error of $\delta z \approx 0.0005$. In other cases, we
could only match the broad blueshifted SN features; the resulting redshifts
($z_{\rm SN}$) vary in quality, with $\delta z \sim 0.005$ (see discussion
below).

In the following sections, we discuss these classification efforts along with
our work to model and remove the host-galaxy contamination and extinction. The
result is a set of ``clean'' supernova spectra with estimates of the quality of
the host subtraction and measurements of the agreement with standard templates
of the best-fitting supernova type. The latter may be used to flag
supernovae with significant spectral anomalies, useful in improving our
understanding of supernova physics and improving our calibration of the
supernova luminosities.

\subsection{Cross-Correlation Analysis}

With a suitable set of templates, cross-correlation analysis provides a
well-established way to measure redshift and to constrain the spectral type.
In the method developed by \cite{to79}, one computes the cross correlation
$c(n)=s(n)*t(n)$ of the original $s$ and template $t$ spectra at zero
redshift, after scaling the wavelength axis of $t(n)$ by a factor $(1+z)$. 
First the spectra are continuum subtracted, $\ln(\lambda)$ binned,
endmasked and filtered to remove any intrinsic color dependency, low-frequency
spectral variations, and high-frequency noise beyond the resolution. The data
are trimmed to a range appropriate to each telescope/spectrograph combination
to avoid the low S/N ends and regions where sky-subtraction errors dominate
(e.g., we truncate HET spectra at 8300 \AA). The code determines the
wavelength shift from a fit to the correlation peak; a measure of the fit
quality comes from $r$, the ratio of peak height $h$ to the root-mean-square
(rms), and $\sigma_a$, of the antisymmetric component of $c(n)$ about the
correlation redshift. For sufficiently large $r$ \citep[typically $>
3$;][]{to79,bl07a,bl07b}, and sufficiently large overlap of $s$ and $t$ in wavelength
\citep[defined as the portion of spectrum used for
cross correlation;][]{bl07a,bl07b}, the measurement of the shift is deemed
significant.

In our analysis we use the ``rvsao.xcsao'' cross-correlation package of IRAF. 
Our supernova templates are from Peter Nugent's spectral
library\footnote{http://supernova.lbl.gov/$\sim$nugent/nugent\_templates.html .}
\citep[see][and the SUSPECT
database\footnote{http://bruford.nhn.ou.edu/$\sim$suspect/index1.html .}; Table
\ref{tbl:tmp}]{nu02}. Specifically for the SNe Ia, we used Nugent's
Branch-normal \citep{br93}, \object{SN 1991T}-like \citep[e.g.,][]{fi92a}, 
and \object{SN 1991bg}-like \citep[e.g.,][]{fi92b} templates,
[supplemented by spectra of some peculiar SNe~Ia (e.g., \object{SN 1999aa}, \citealt{gara04}; \object{SN 1999by}, \citealt{garn04}, etc.) from SUSPECT. For SNe Ib/c, the templates 
used are Nugent's normal SN~Ib/c and hypernova spectra, as well as observed spectra of
\object{SN 1990B} \citep{ma01}, \object{SN 1990I} \citep{el04}, \object{SN 1994I} \citep{cl96}, 
\object{SN 1998bw} \citep{fe01}, \object{SN 1999ex} \citep{ha02}, \object{SN 2000H} \citep{br02}, and \object{SN 2002ap} \citep{ga02}, from the SUSPECT database]. 
Similarly, the set of Nugent's SN~II-P, SN~II-L, and SN~IIn spectra,
augmented by SUSPECT's SN~II and peculiar SN~II spectra, constitutes the template
library for SNe II. These templates represent a variety of supernova ages.
In particular, the SN~Ia templates cover a wide range of epochs ($-19$ to 70~d
from peak $B$-band magnitude). Galaxy templates were drawn from the SDSS
catalog\footnote{http://www.sdss.org/dr5/algorithms/spectemplates/index.html .}
and covered five major morphological types: E/S0, Sa, Sb, Sbc/Sc, and Sm/Im.

\begin{deluxetable}{cc}
\tabletypesize{\scriptsize}
\tablecaption{Cross-Correlation Templates From SUSPECT SN Database \label{tbl:tmp}}
\tablewidth{0pt}
\tablehead{\colhead{Type} & \colhead{Object} }
\startdata
Ibc    &    1954A, 1962L, 1964L, 1990B, 1990I, 1994I, 1998bw, 1999ex, 2000H, 2002ap    \\
II    &    1948B, 1959D, 1961F, 1961I, 1961V, 1964F, 1964H, 1969L, 1970A, 1986E    \\
II    &    1987A, 1988A, 1988H, 1988Z, 1989C, 1990ae, 1990ag, 1990E, 1990H, 1990K \\
II    &    1990Q, 1990V, 1990X, 1991C, 1991H, 1991J, 1992aa, 1992ab, 1992C, 1992H    \\
II    &    1993J, 1994aj, 1996L, 1997ab, 1997cy, 1997D, 1998dn, 1998S, 1999em, 2004dj    \\
Ia-pec    &    1957A, 1960H, 1986G, 1991bg, 1991bj, 1991F, 1991T, 1997br, 1997cn, 1999aa \\
Ia-pec    &    1999ac, 1999by, 2000cx    \\
\enddata
\end{deluxetable}

When fitting for galaxy redshifts $z_{gal}$, we adjust the xcsao Fourier
filtering to focus on narrow features (high frequency in Fourier space). The
Fourier cutoff is at lower frequency when matching SN templates to derive
$z_{\rm SN}$. The cross-correlation program provides a best-fit redshift, $r$
value, and overlap wavelength range for each template in our library. The
relative value of $r$ at the best-fit redshift provides a guide to the best
template class. We discard results with less than 40\% of the spectra in the
overlap region \citep{bl07a,bl07b}. All measurements are made in the heliocentric
frame.

We apply the cross-correlation analysis above to every spectrum obtained and
inspect the templates with highest $r$ values and adequate spectral overlap.
As expected, the redshift solutions are very insensitive to the template; for
example, fits with different galaxy types, such as Sb and Sm/Im, give $\delta
z \lesssim 0.001$. Supernova fits are of course somewhat less
precise. Nevertheless, comparing a set of SN Ia and SN Ibc fits gives $\delta
z \lesssim 0.005$.  Thus, the $z$ measurements reported here are robust.  One
point deserves comment: half of our spectra have both SN and host-galaxy
redshift solutions, and there is a slight, but statistically significant,
systematic offset between the two redshifts (see Figure~\ref{fig:zgzsn}).
The difference amounts to $\delta z \sim 0.003$ or a velocity shift of $\sim
900~\rm{km~s}^{-1}$, which is only a small fraction of the photospheric
expansion velocity of the SN ($v \sim 10,000~\rm{km~s}^{-1}$ near maximum
light).  Different SNe have different expansion velocities (see, e.g.,
\citealt{be05}) and temporal evolution.  Since there is no obvious systematic
mis-match between the spectroscopic and photometric epochs
(Figure~\ref{fig:epoch_pc}), we believe the difference is most likely due to
the fact that the average photospheric expansion velocity observed in our SN
sample is slightly different from that assumed in Nugent's Branch-normal SN~Ia
templates, which is strictly valid for SN~Ia with a stretch factor of unity.
When only a supernova redshift is available, we correct the $z_{\rm SN}$ for
this systematic offset 0.003 (the mean of Figure \ref{fig:zgzsn}). However,
when a significant $r$ is obtained for a galaxy fit $z_{gal}$, we adopt these
more accurate redshifts. The dispersion in the $z_{\rm SN}$/$z_{gal}$
comparison of $0.005$ is also adopted here for the systematic error in the
$z_{\rm SN}$ measurement. This error is generally larger than the statistical
error ($\delta z \approx 0.001$) associated with rvsao.xcsao fit results, as
it includes systematics associated with uncertainty in the SN velocity and
epoch. We compared our galaxy measurements with redshifts from the SDSS
database, where available (about a fifth of our sources, primarily at low
redshift), finding agreement to within $\delta z=0.001$; no systematic offset
is observed.

\begin{figure}[t]
\begin{center}
\begin{tabular}{cccc}
  \epsfig{file=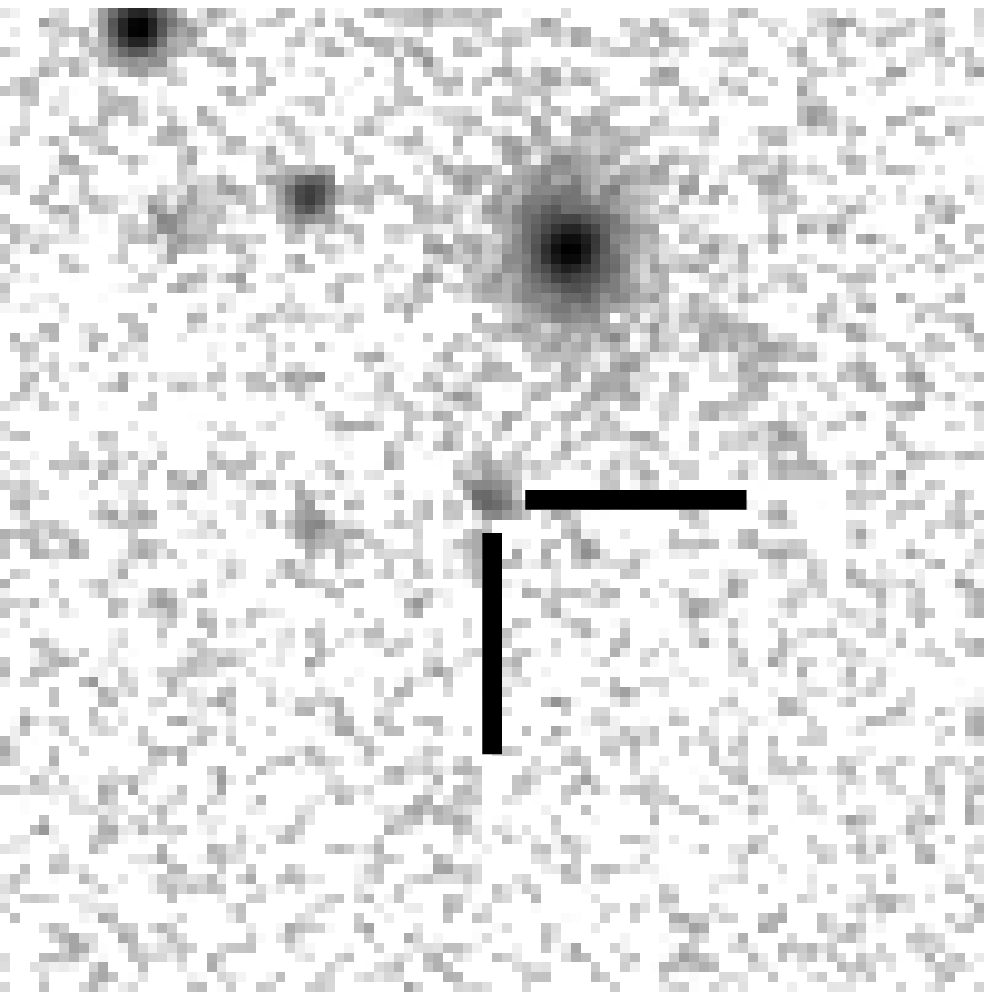, width=1.5in,clip=} &
  \epsfig{file=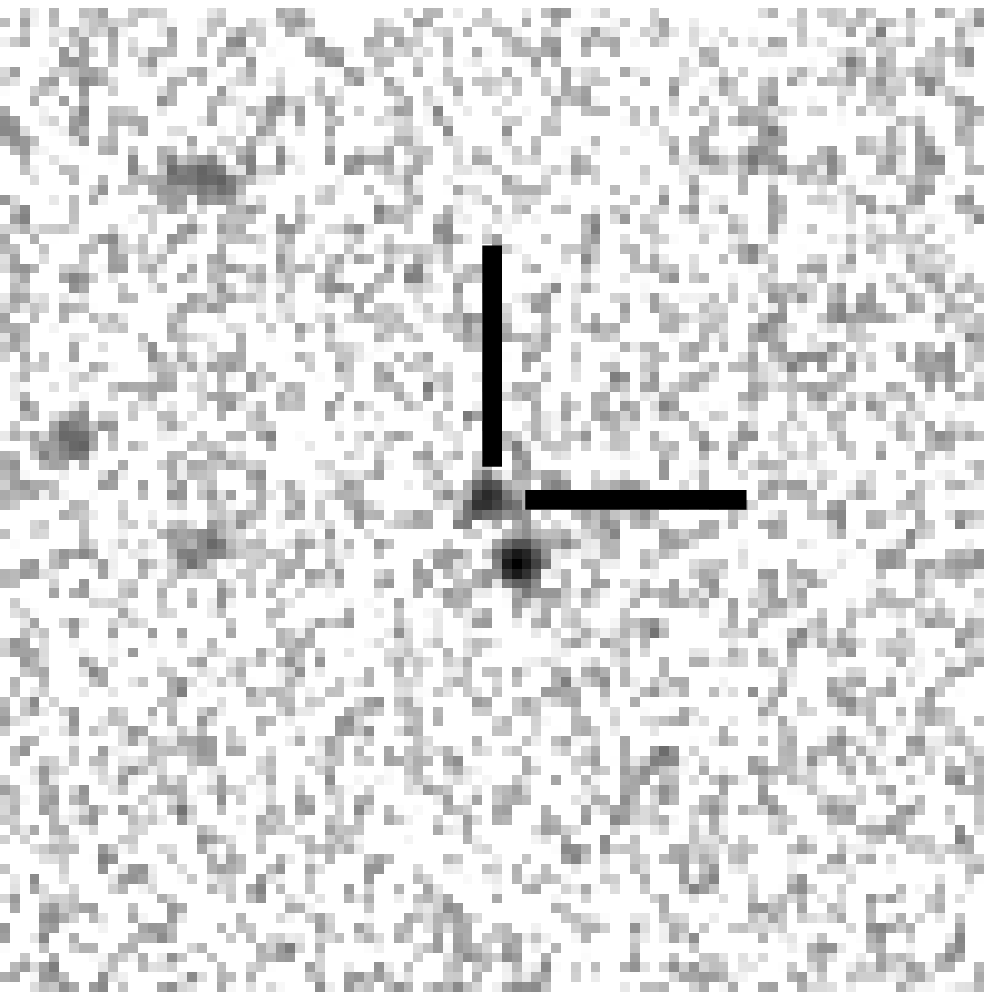, width=1.5in,clip=} &
  \epsfig{file=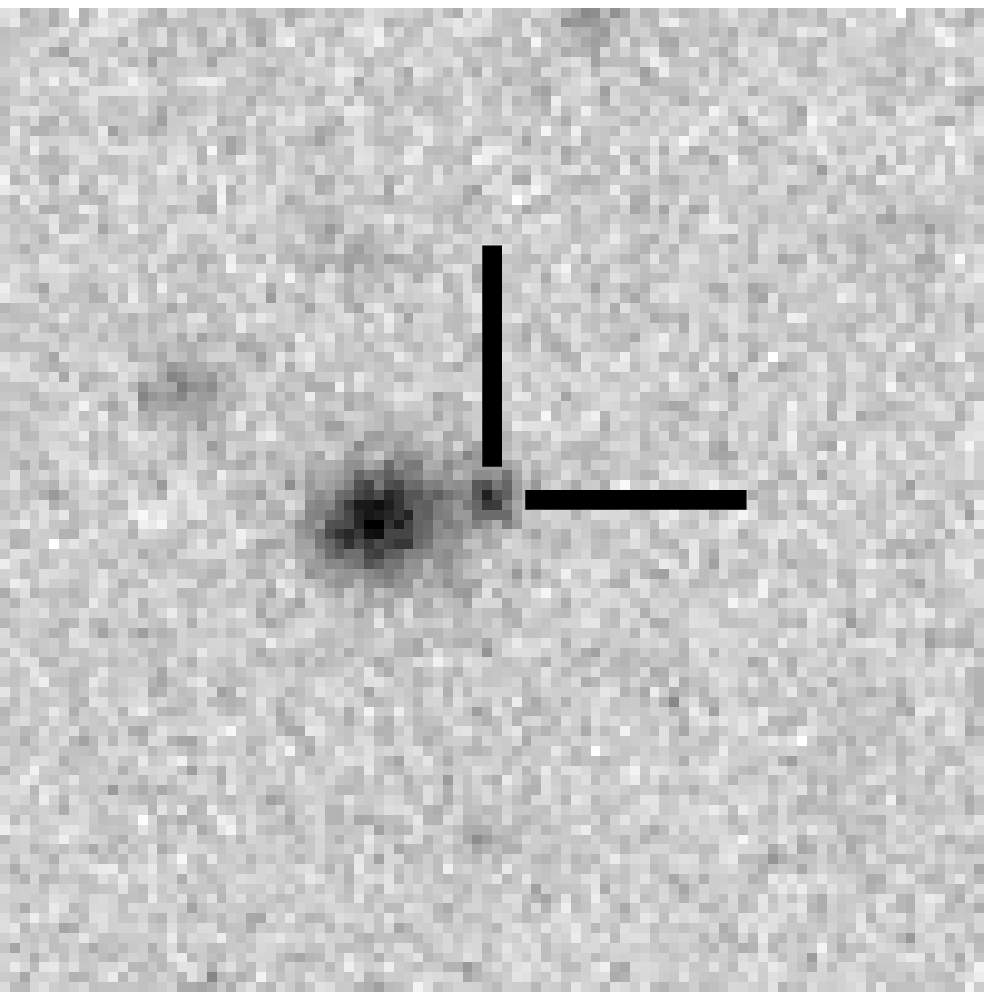, width=1.5in,clip=} &
  \epsfig{file=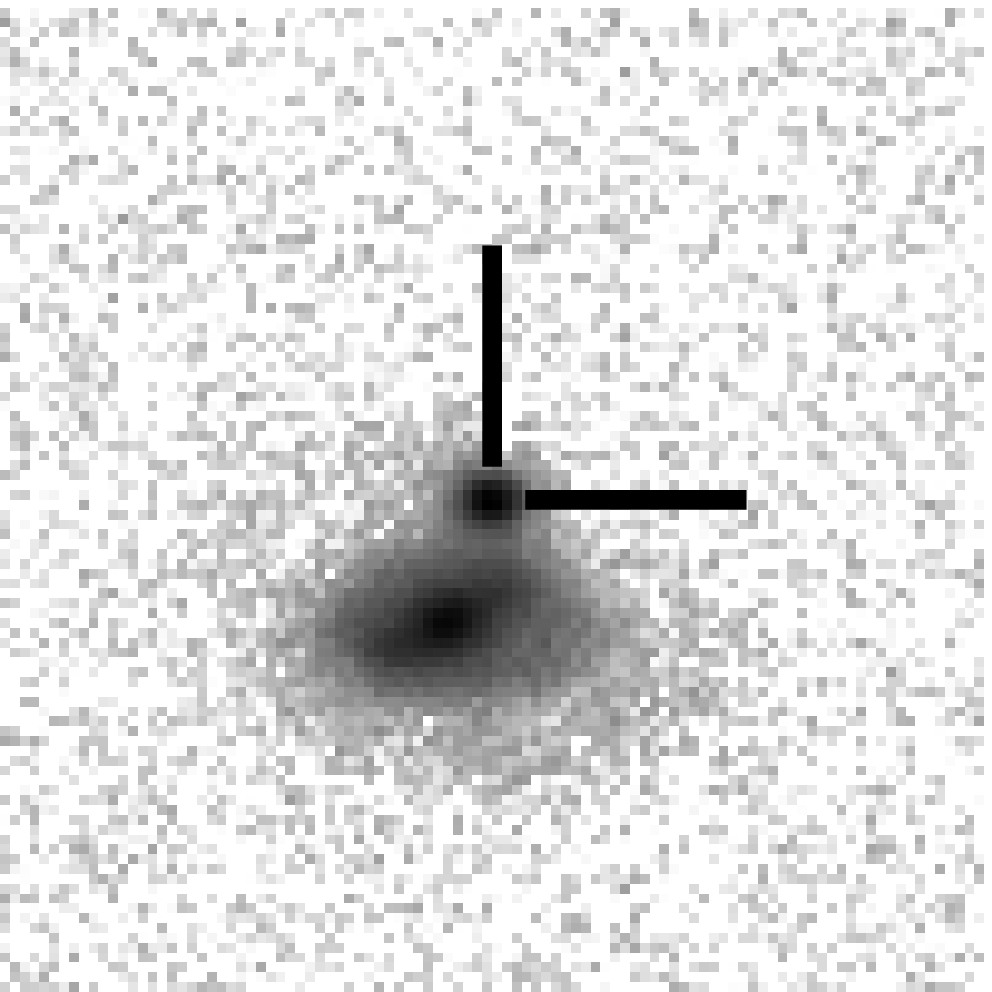,width=1.5in,clip=} \\
 \\
  \epsfig{file=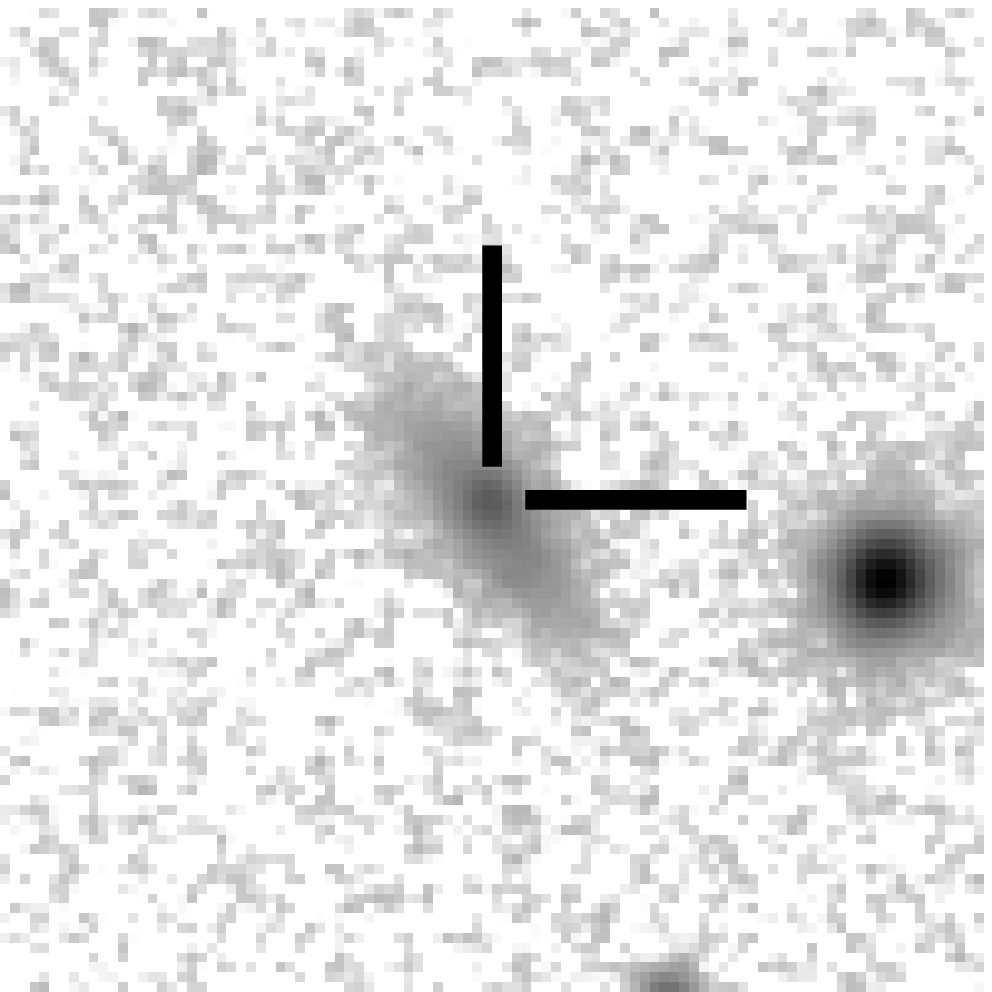, width=1.5in,clip=} &
  \epsfig{file=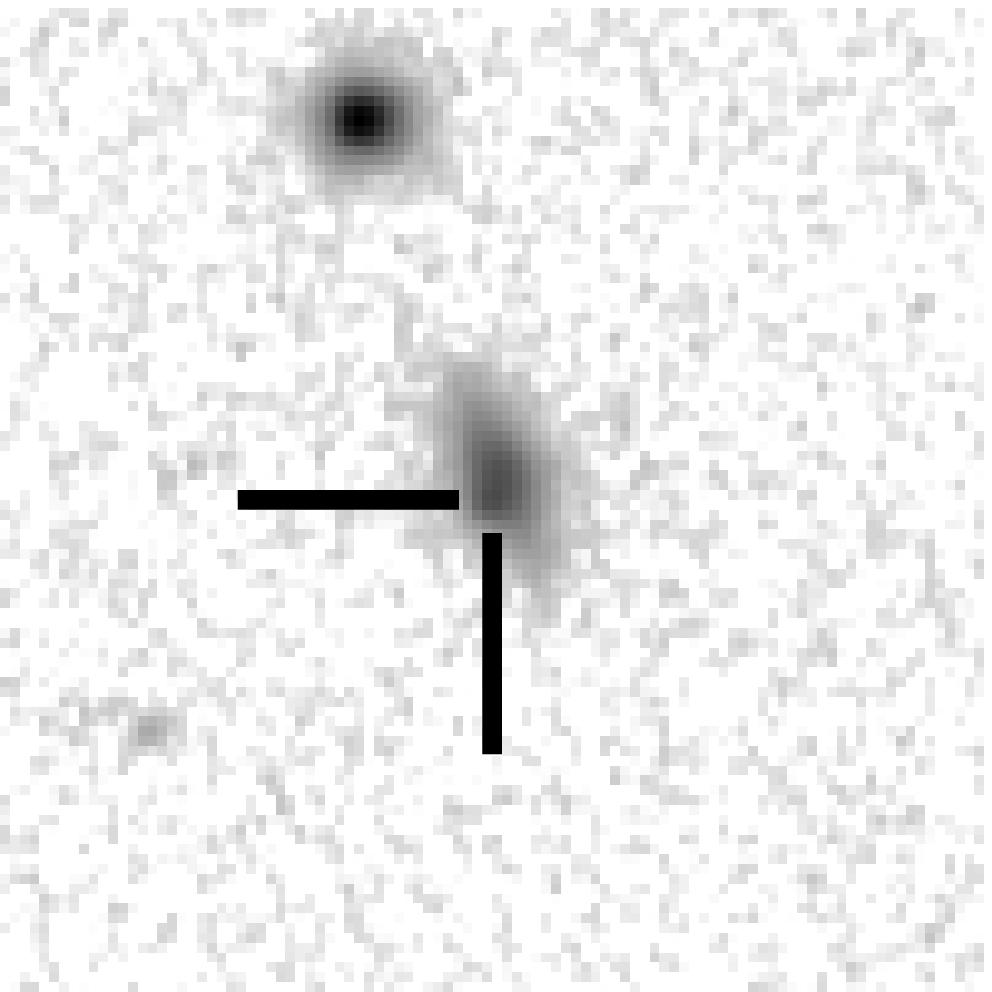, width=1.5in,clip=} &
  \epsfig{file=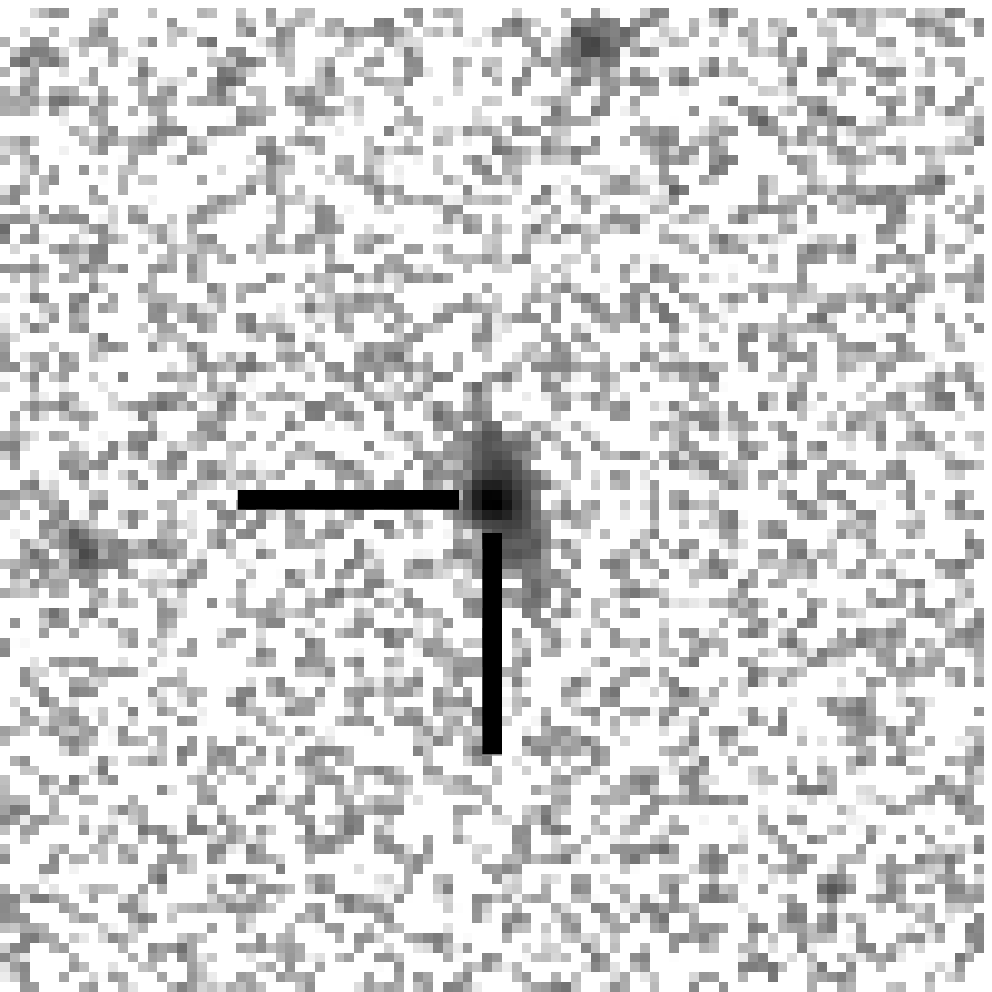, width=1.5in,clip=} &
  \epsfig{file=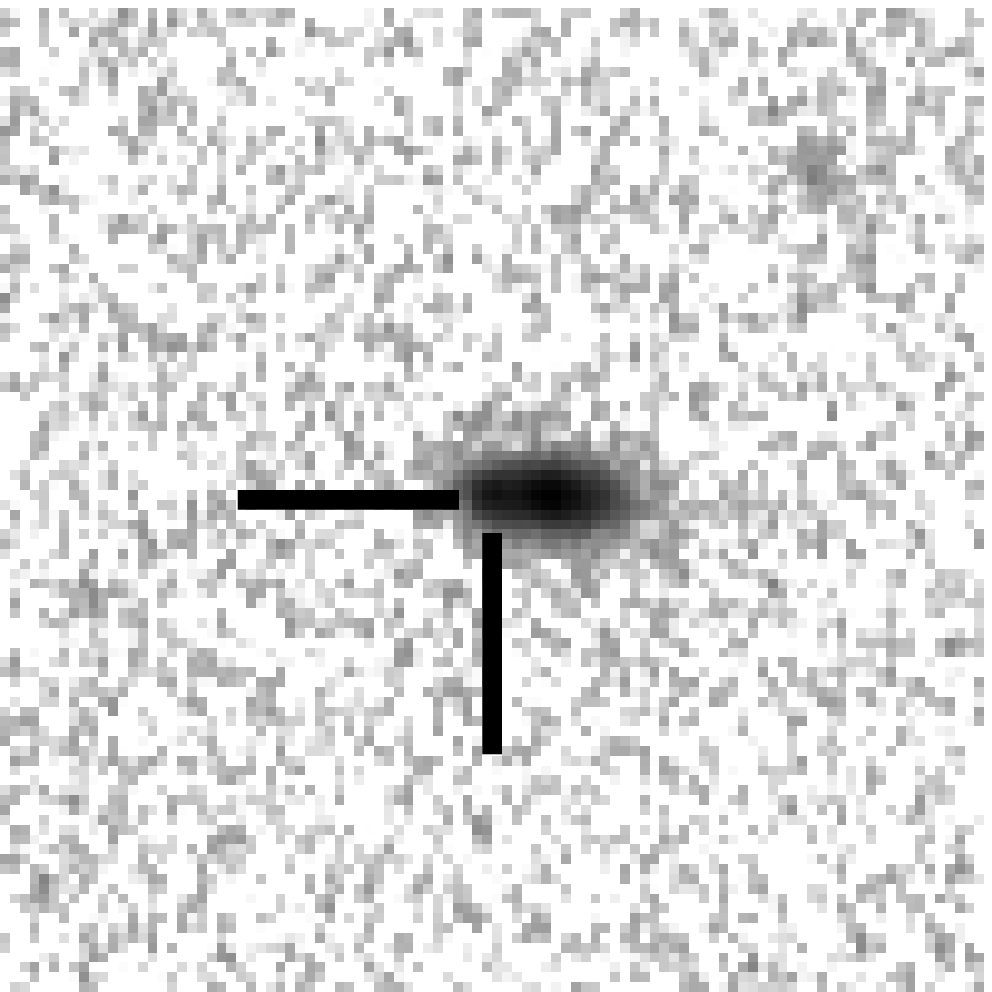, width=1.5in,clip=} 
\end{tabular}
\caption{This panel shows images of ``clean" (top row) and 
``heavily contaminated" (bottom row) supernovae. The tick marks indicate the 
location of the supernova. \label{fig:imag}}
\end{center}
\end{figure}

Comparison of $r$ values between different templates can be used to select the
most likely spectral class. For high-quality spectra with little or no
host-galaxy contamination, the $r$ values usually cleanly select one SN
type. However, for other spectra, we had to visually inspect the solutions at
the local $r$ maximum for each SN/galaxy type. This was done by plotting the
data over the combined redshifted SN/galaxy spectrum, with a visual
normalization, and inspecting how well the features match. Generally, the
largest $r$ value was indeed preferred, but in a few cases inspection showed
that the largest $r$ corresponded to a misleading fluctuation in the template
matches.  High-redshift and low-quality spectra, in particular, often showed
multiple local minima and human judgment was needed to select the best
match. For a small fraction (less than 10\%) of our spectra that have very low
S/N ratio, none of these local minima makes a reasonable fit and thus the
spectra were classified as ``unknown" and not included in this paper.

We found that $r$ did not select as well between host-galaxy classes as
between SN types. A clear differentiation between absorption-line dominated Sa
and emission-line dominated (Irr) spectra is available whenever the S/N and
host contamination are fairly large. However, finer distinctions (e.g., Sa/Sb)
were not reliable. Indeed, for our multi-epoch spectra, $\sim$40\% of the time
the best-fit host template differs by a class between epochs.

We spectroscopically identified SNe Ia according to the following criteria:
(1) $r \ge 3$ and (2) a high-significance detection of at least one Si II
absorption-line feature (at rest wavelengths of $\sim$ 4000 \AA, $\sim$ 5800
\AA, and $\sim$ 6150 \AA). For lower values ($r \approx 3$),
we additionally required a spectrum/template spectral overlap $> $60\% (lower
overlap values usually indicate a poor type and redshift range) and/or a host
with a spectroscopic classification of E/S0. A requirement of $r \ge 3$ for
SN~Ia identification has also been adopted by other groups
\citep{bl07a,bl07b,ma05}. When these criteria are satisfied, we designate the
SN as Type Ia (``Ia''). When one or more of these criteria fail, but the best
match is still a Ia template, we mark the object as ``Ia?'' -- a
spectroscopically probable SN~Ia. In particular, for late-time spectra the $r$
discrimination was often poor between Types Ia and Ib/c. Occasionally, both
SN~Ia and SN~Ia-pec types gave acceptable fits. In these cases, host-galaxy
type, absolute magnitude, and the extent of the spectral overlap were helpful
in making the final classification. Thus, while the redshift estimates are
quantitative and, for a given spectrum type, choice of the best-fit SN age is
quantitative, the selection of the best type match is necessarily somewhat
subjective.

Once a type is confirmed, we find that the $r$ values resolve the supernova
phase to an accuracy of $\sim 2$ to $3$~d, similar to other SN
template-fitting techniques \citep{ba06,bl07a,bl07b}. The basic results from
the cross-correlation analysis are listed in Table \ref{tbl:spec}.

\subsection{Spectral Decomposition and Host-Galaxy Subtraction}

The majority of our SN spectra have substantial host-galaxy contamination
(Figure \ref{fig:hostcontam} and Figure \ref{fig:imag}), because of the wide
slits used and relatively high redshift and small angular size of the
hosts. Indeed, for late-time SN spectra and subluminous objects, the
host-galaxy light coincident with the supernova often dominates that of the
supernova itself. To study SN spectral diversity and to explore the
correlation of spectral parameters with our high-quality light curves, we wish
to have SN spectra with minimum contamination.  Accordingly, we have attempted
to quantitatively measure the SN and host contributions using principal
component analysis (PCA).

\begin{figure}
\begin{center}
\includegraphics[angle=0,scale=.50]{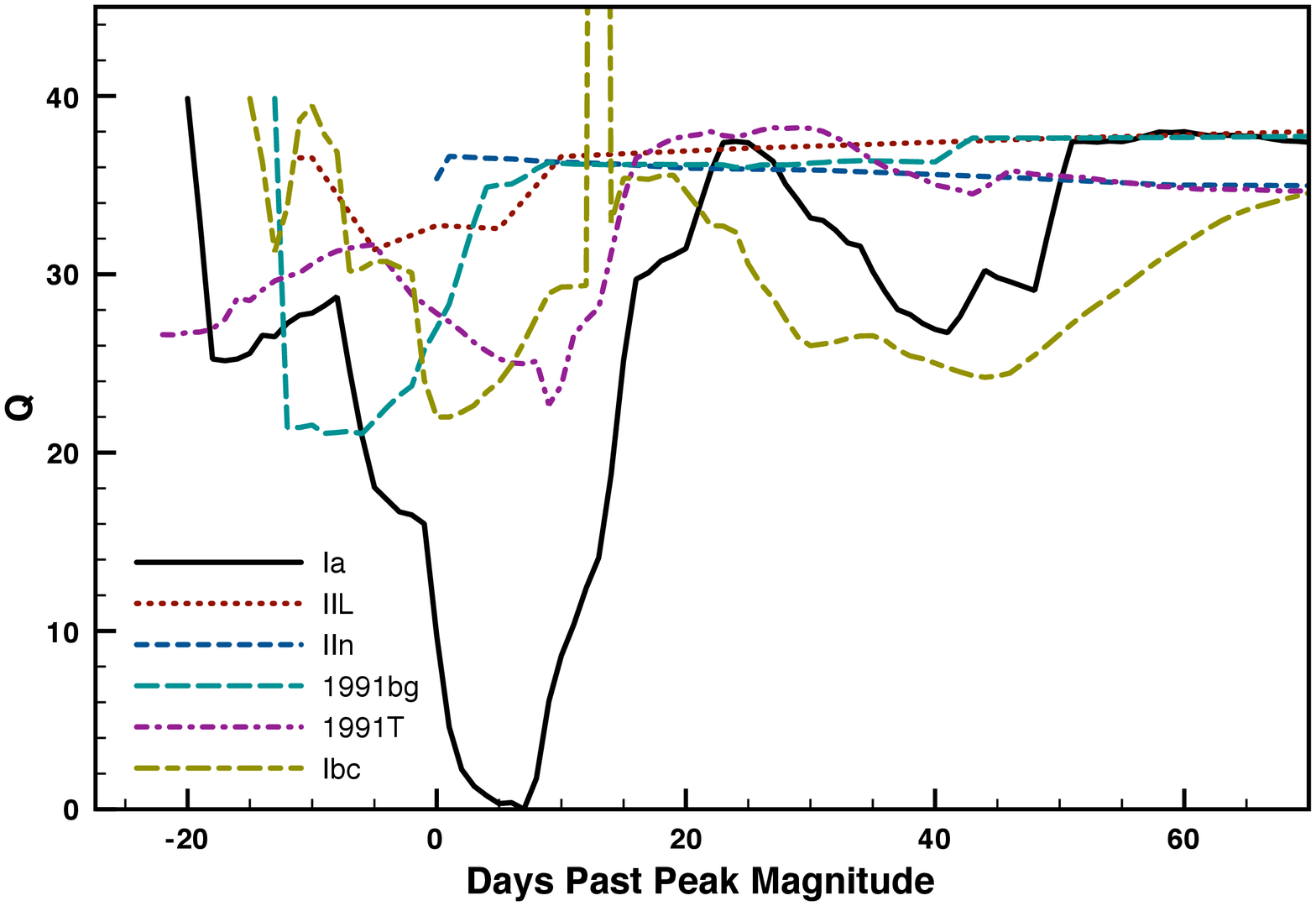}
\includegraphics[angle=0,scale=.50]{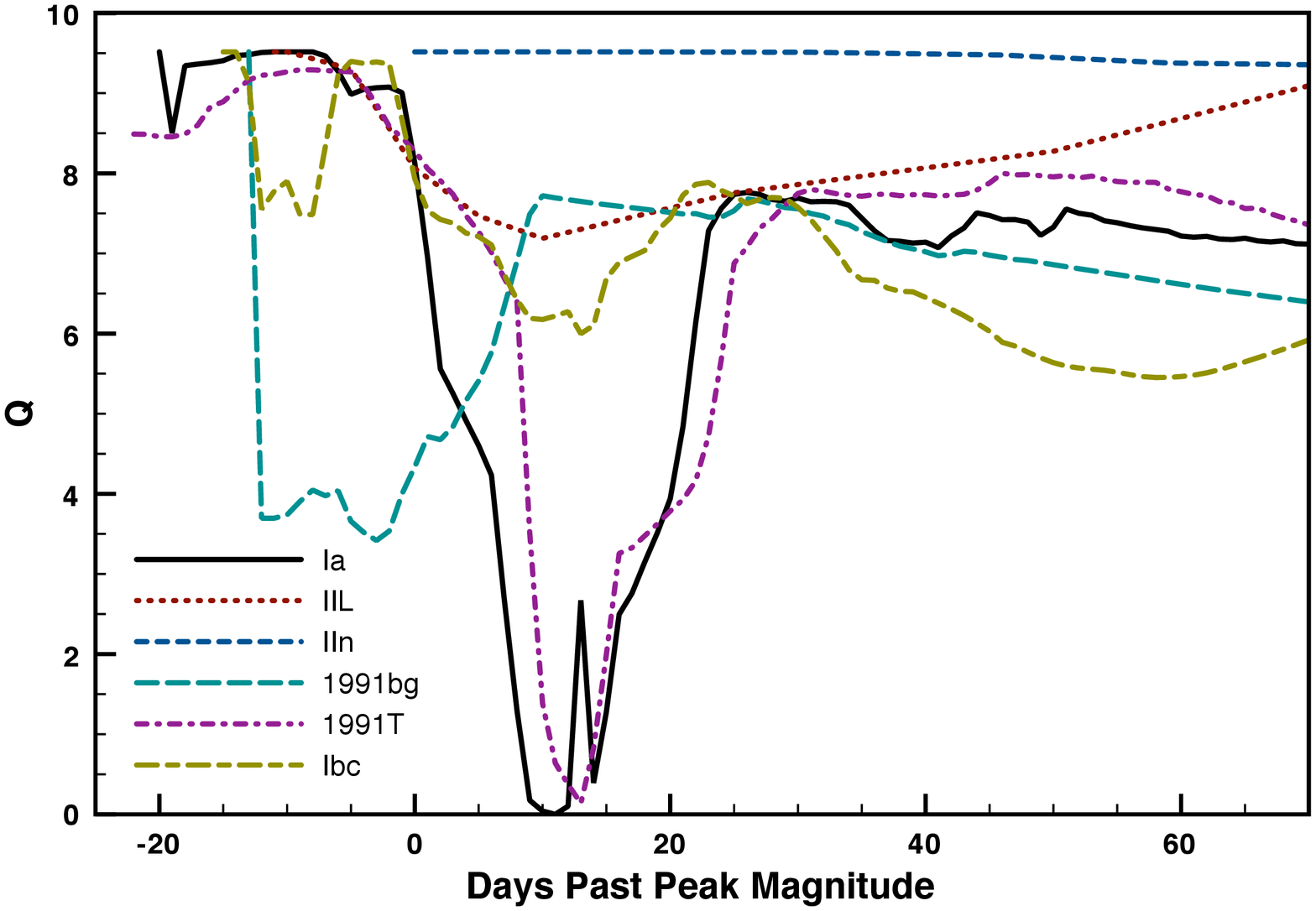}
\caption{Two panels showing $\chi^2_{\nu}$ fitting results to various SN
  templates as a function of age. The object in the upper panel, with $\nu =
  943$ degrees of freedom, is classified as a SN~Ia at high confidence, since
  the quality factor $Q = \Delta\chi^2_{\nu}\sqrt{\nu}/\chi^2_{\nu}$ of the
  next-best type solution is $\gg$1. The lower panel, with $\nu = 918$ degrees
  of freedom, has a preferred but not definitive SN~Ia classification, with $Q
  \ll 1$. The photometric ages for these two objects are 5~d and 16~d after
  $B$-band maximum respectively; the spectroscopic age estimates are
  consistent with this value.
\label{fig:chi}}
\end{center}
\end{figure}

Several approaches have been developed for such spectra decomposition. One
method \citep{mad03} projects the observed spectrum onto the
Karhunen-Lo\`{e}ve (KL) transformation or PCA eigenspectra basis \citep{co95,
  fo96, ma02} to extract the reconstructed source (e.g., galaxy) spectrum.
Other techniques use point-source deconvolution \citep{bl05} or $\chi^2$
template-fitting techniques \citep{ho02, li05, hoo05, ho05}, modeling both SN
and galaxy with template libraries. However, this latter approach sometimes
fails, yielding only a low S/N detection of the SN spectrum. In such cases, it
is common to resort to a cross-correlation technique similar to that used
above \citep{ma05}.

Here we adopt a composite PCA + template-fitting program to separate the SN
spectrum from the host-galaxy light. In this procedure, each observed spectrum
is modeled by a weighted combination of SN templates (identical to those used
for the cross-correlations analysis) and PCA eigenspectra derived from the
SDSS galaxy sample \citep{yi04}. The redshift is fixed to the value determined
by the cross-correlation procedure above, but we allow the flux from the
supernova and the three dominant PCA galaxy components to vary. In addition,
we optionally add a variable amount of absorption with a Galactic extinction
law $A_\lambda$ \citep[standard reddening,][and $R_V=3.1$]{ca89} at the host
redshift. To monitor the fit we define a Figure of Merit (FoM) normalized by
the degrees of freedom,
\begin{equation}
\chi^2_{\nu} = \sum_{\lambda} {\frac{[D(\lambda)-a S(\lambda,T) 10^{c
    A_{\lambda}} \ - \sum_{i=0}^{2} {b_i G_i(\lambda)}]^2}{\nu
  (\sigma(\lambda))^2}},
\end{equation}
where $D$ is the observed host + SN spectrum, $S$ is the SN template spectrum
at an epoch $T$, $G_0$, $G_1$, and $G_2$ are the first three major PCA
eigenspectra of the SDSS galaxy sample \citep{yi04}, $A_{\lambda}$ is the
extinction (applied at the redshift of the SN), and $\sigma$ is the 1$\sigma$
error of the spectrum. Here $a$, $b_0$, $b_1$, $b_2$, and $c$ are constants to
be adjusted for the best fit in the SN spectra, host galaxy, and reddening
space. The values of $b_0$, $b_1$, and $b_2$ indicate the galaxy type of the
host \citep{yi04}. We neglect extinction and reddening in the Milky Way, which
produces $E(B-V) < 0.1$ mag for almost all of the survey stripe.  Here the
$\nu$ degrees of freedom are determined as $N-m$ where $N$ is the number of
data points and $m$ accounts for the fitting parameters and the (nearly
continuous) date parameter (time from maximum light). Since the first three
eigencomponents contain $98.2\%$ of the total SDSS galaxy sample variance, the
amplitudes $b_{0,1,2}$ can classify $\sim 99\%$ of observed galaxies
\citep{yi04}. These components thus provide a reasonable representation of the
host spectrum, avoiding unphysical galaxy models which can be derived when
many low significance eigencomponents are included.

\begin{figure}
\begin{center}
\includegraphics[angle=0,scale=.5]{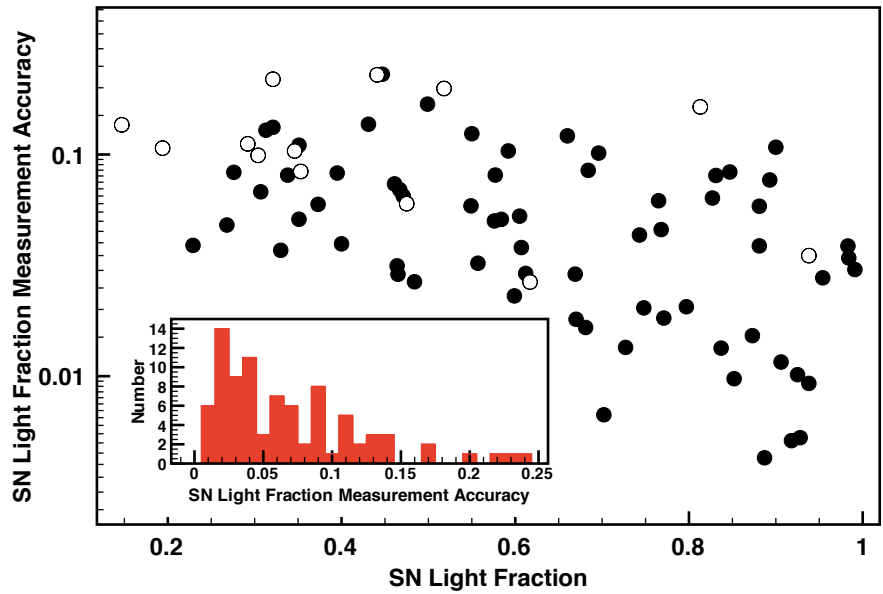}
\caption{The accuracy of the SN component measurement (defined as the
  statistical error of SN light flux divided by the fraction of SN light flux)
  plotted vs. SN fraction. ``Type Ia?'' (empty circles) tend to have
  relatively low accuracy. In most cases, the SN fraction is measured to
  within $<10 \%$. \label{fig:sigmasn}}
\end{center}
\end{figure}

\begin{figure}
\begin{center}
\includegraphics[angle=0,scale=.5]{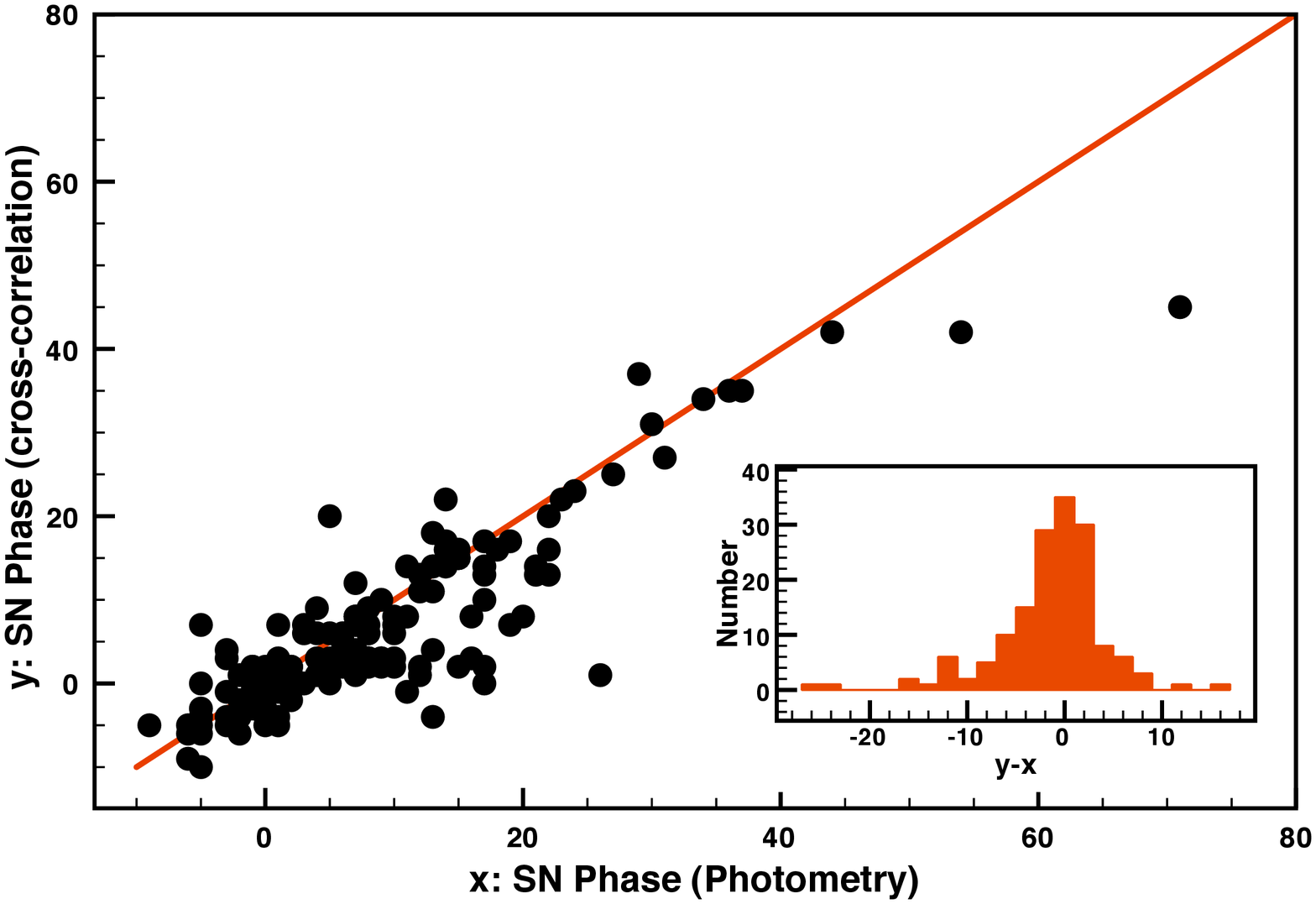}
\includegraphics[angle=0,scale=.5]{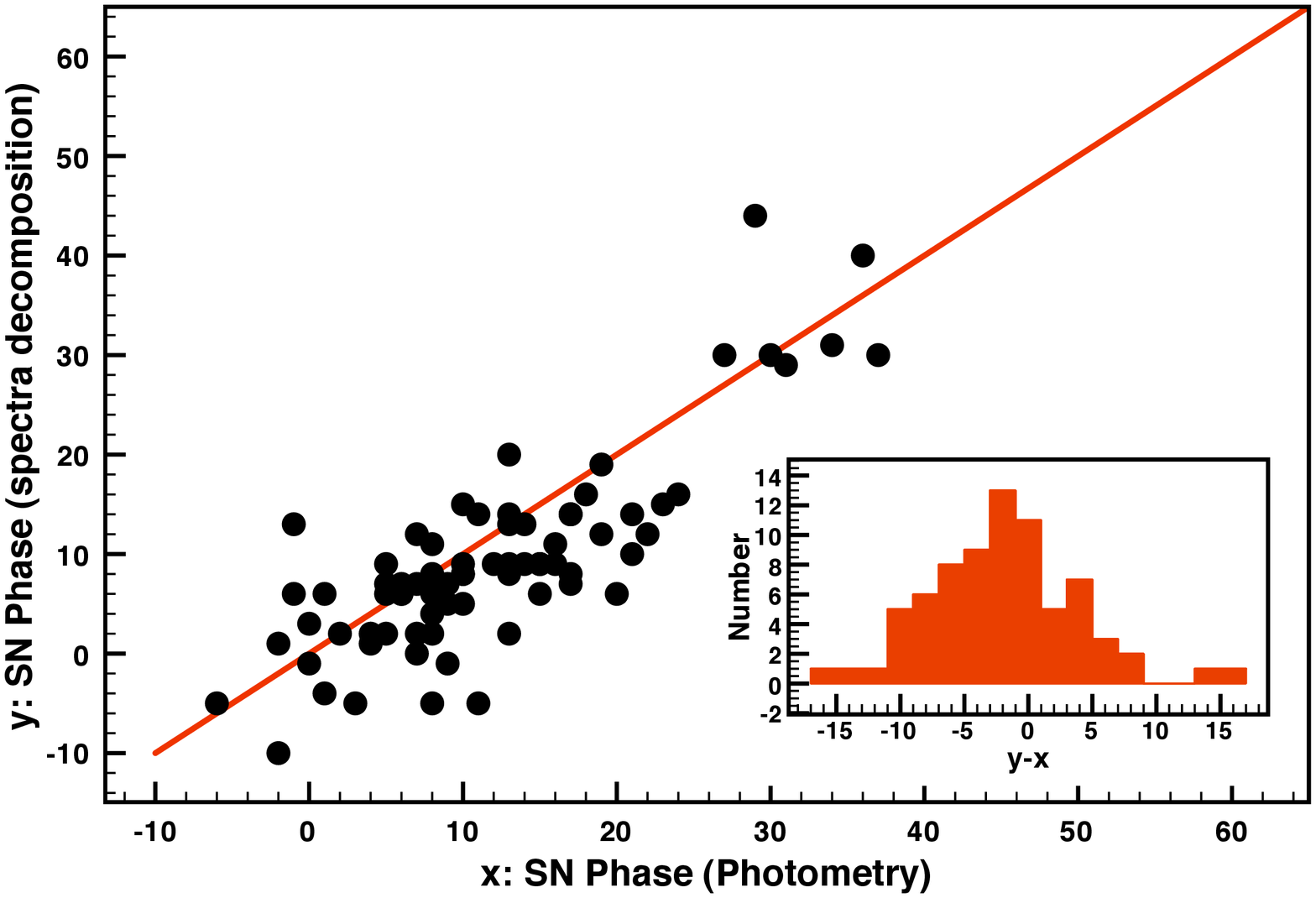}
\caption{Two panels comparing different-epoch measurements of confirmed
  SNe~Ia. The upper panel shows the SN phase (days past peak $B$-band
  magnitude) measured from cross-correlation analysis of the spectra against
  the photometric phase. The lower plot shows the SN phase measured from the
  PCA + template-fitting program against the photometric phase. The insets
  show histograms of the scatter between the two epoch measurements, with $y$
  being the phase determined from either cross-correlation analysis or the
  spectra decomposition analysis and $x$ being the phase determined from
  light-curve fitting. \label{fig:epoch_pc}}
\end{center}
\end{figure}

\begin{figure}
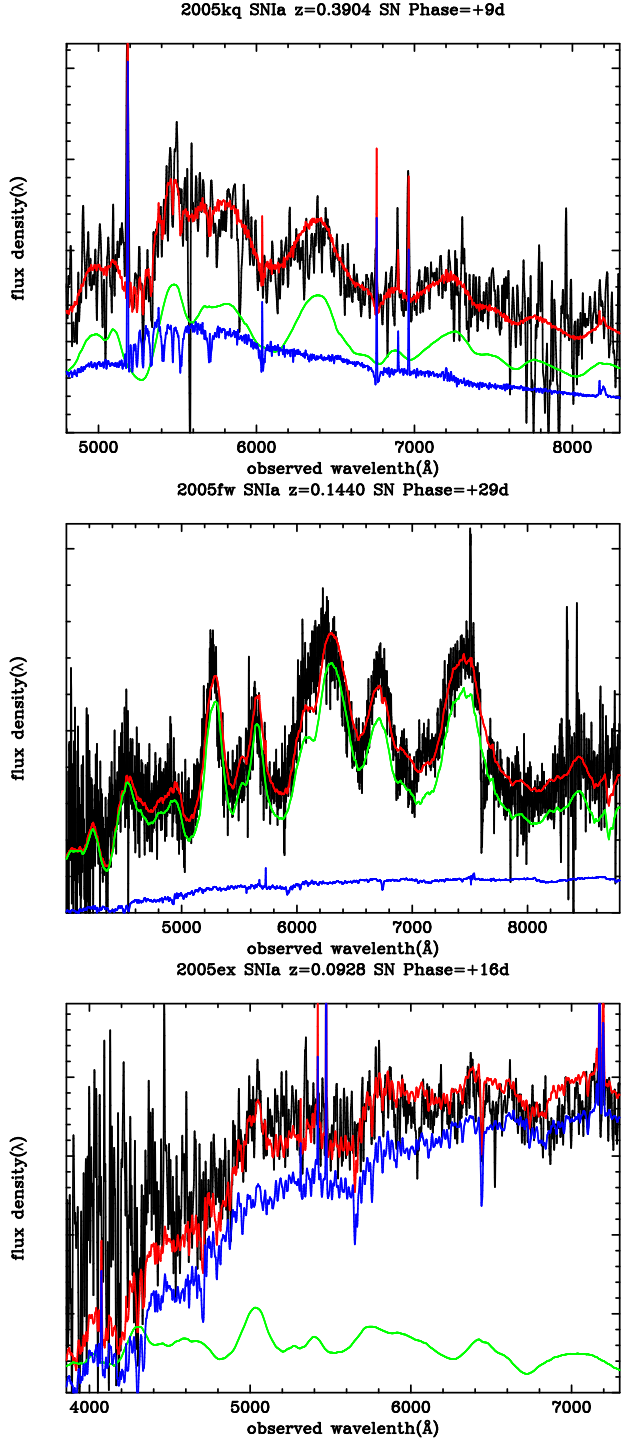

\begin{center}
 \begin{minipage}[t]{0.35\linewidth}
  \epsfig{file=HET.eps, width=2.5in, angle=270} \\
  \epsfig{file=Sub.eps, width=2.5in, angle=270} \\
  \epsfig{file=MDM.eps, width=2.5in, angle=270} 
  \end{minipage} \hfill
  \begin{minipage}[t]{0.42\linewidth}
     \caption{Examples of PCA spectral decomposition. The data (black) are
       overplotted with the best-fit combined model (red), with host-galaxy
       (blue) and supernova (green) components. The header of each figure
       shows the name, redshift, and phase of thesupernova. Wavelength is
       shown in the observed frame. The HET spectrum (upper left, resolution:
       20 \AA) has almost an equal amount of supernova and host light. The
       Subaru spectrum (middle left, resolution: 8--12~\AA) has little host
       contamination, while the MDMspectrum (bottom left, resolution: 15 \AA)
       is dominated by the host. \label{fig:exam}}
    \end{minipage}    
\end{center}  
\end{figure}

This procedure gives the best decomposition for a given SN template spectrum
$S$. We can, of course, then perform this minimization for each plausible $S$
(i.e., different supernova types and ages). Figure \ref{fig:chi} shows two
examples of how the figure of merit $\chi^2_{\nu}$ varies as $S$, type, and
date change. In the upper panel of Figure \ref{fig:chi}, the large improvement
near the best global maximum shows a definitive typing as a SN~Ia (and a
well-constrained date and decomposition). In the bottom panel of
Figure~\ref{fig:chi}, a standard SN~Ia also provides the best fit, although a
SN~1991T template is only modestly worse. Note, however, that both acceptable
types show a similar well-constrained age.

If the data had Gaussian errors with the correct amplitude and the model
describes the data well, we would expect $\chi^2_{\nu}$ to approach 1, with a
spread of $\Delta \chi^2_{\nu} \approx \nu^{-1/2}$ where $\Delta \chi^2_{\nu}$
stands for the difference between one solution and the best-fit solution. In
practice, however, the errors may be mis-estimated and/or the model may be
wrong, giving larger $\chi^2_{\nu}$. One quantity that can be used under these
conditions to distinguish between two models is a quality factor defined by
\begin{equation}    
Q = \frac{\Delta \chi^2_{\nu} \sqrt{\nu} }{\chi^2_{\nu}},
\end{equation}
where the $\chi^2_{\nu}$ value for the best-fit model is inserted in the
denominator to compensate for incorrect error estimates. Large $\Delta
\chi^2_{\nu}$ yields large $Q$ values and thus more significant discrimination
between the models. In fact, even when the models are not correct or equivalently the data suffer
uncorrected systematic bias, a large value of $Q$ ($\ga 1$) can indicate a
significant discrimination between two models.

Unfortunately, we find that we do not always satisfy this criterion ($Q \ga
1$), even when visual inspection indicates a clear type assignment, so our PCA
analysis does not always definitively type the SN. While it does indeed
confirm the clear SN identifications, and while the ``best'' type (albeit at
small $\Delta \chi^2_{\nu}/\chi^2_{\nu}$) is often that selected by hand,
there are some disagreements with the types selected manually from the $r$
values of the cross-correlation analysis. At least for our typical data
quality, it seems that PCA analysis does not provide a reliable automatic
classifier for low-S/N spectra. In particular, the $\Delta
\chi^2_{\nu}/\chi^2_{\nu}$ analysis has difficulty distinguishing between
SN~Ia subtypes. Thus, in ambiguous cases, we defer to the typing from the
cross-correlation analysis. However, even in these cases, the PCA analysis is
valuable, as it provides a quantitative estimate of the galaxy/SN
decomposition.

As noted above, we can, in principle, independently measure the absorption
from the spectrum reddening.  However, when we allow $c$, the coefficient that
characterizes the amount of extinction and reddening, to range freely, the
value at $\chi^2_{\nu}$ minimum is often higher than that determined from the
light-curve analysis.  One difference is that the light-curve analysis
includes an exponential prior (favoring low $A_V$ values), while we do not
impose a prior.  However, we believe that the limited accuracy of our relative
spectrophotometry prevents us from making useful reddening measurements.  This
is seen in the large dispersion of the fit values.
Also it may be problematic that our assumed extiction law, which corresponds
to the standard Galactic $R_V=3.1$, may not apply to SNe Ia host environments.
Indeed widely values have been estimated for $R_V$ (e.g. \citealt{br92}; 1.55,
\citealt{kr06}; 2.5, \citealt{kn03}, \citealt{al04}, etc.). 
Accordingly, for our final decomposition we fix the value of the extinction to
that obtained from the light-curve analysis.  The redshift and spectral
identification are not very sensitive to the value of $A_V$ to within their
uncertainties.

Thus, adopting the $z$ and SN type from the cross-correlation analysis and the
extinction from the light-curve fitting, we can use the PCA decomposition to
obtain a model galaxy spectrum and a fraction of SN light together with
statistical errors (two-sided, asymmetric error bars with a 1$\sigma$
confidence level). The accuracy of such measurement vs. the magnitude of the
fraction is shown in Figure ~\ref{fig:sigmasn}. The relatively low
significance points include SNe of the ``Ia?'' class, for which the
characteristic lines were not measured with high significance, but the general
spectral shape was best fit by a SN~Ia template.

In general, the typing from cross-correlation analysis and PCA decomposition
were consistent. The epochs determined from both analysis mostly agree within
$\sim 5 $~days and are close to those estimated from photometry analysis (see
Figure \ref{fig:epoch_pc}). Approximately $80\%$ of the galaxy types
determined from the PCA analyses are consistent with the cross-correlation
results. No prior is assumed except for the physical constraints on the
parameters ($a > 0$, constraints on the $b_0$, $b_1$ and $b_2$ to make sure
the constructed galaxy type is physically plausible). The fit values are
listed in Table~\ref{tbl:host}. Figure \ref{fig:exam} presents typical
synthetic SN/host spectra, with relative amplitudes determined from the PCA
analysis.

\section{CONCLUSION AND FUTURE WORK}

We have presented the follow-up spectroscopy for the Fall 2005 season of the
SDSS-II Supernova Survey. A semi-quantitative procedure of SN identification
is developed, based on cross-correlation techniques. The rvsao.xcsao $r$ value
and the overlap of template and data in wavelength space are useful guides to
the quality of the fit. Nevertheless, a fair amount of human judgment is
required to flag solutions associated with false local minima for low S/N
spectra. Using this procedure, we have determined accurate redshifts from both
SN spectra and host spectra within $\delta z_{\rm SN} \approx 0.005$ and
$\delta z_{gal} \approx 0.0005$, respectively. The typical uncertainty in
phase is $\sim 3$~d.

We have also described our efforts to quantify host-galaxy contamination using a
combined $\chi^2$ fitting and PCA analysis, which provides an efficient way to
give useful decompositions into SN and host-galaxy spectra with $<10 \%$ accuracy,
given known SN types from cross-correlation analysis and host-galaxy extinction 
estimated from the multi-band light curves.
Typically, when the spectrum contains more than
$60$\% host-galaxy light, the host type is well constrained. Both
spectroscopic analyses show good agreement in estimating the photometric
epoch.

Using our quantitative measurement of the SN light fraction, we can make
galaxy-subtracted SN spectra with estimates of the residual galaxy
contamination. With a $> 50$\% larger yield from the second season, 2006,
and an anticipated large number of SNe in the third and final season, the
classified and host-subtracted spectra will provide a large and uniform data
set with which to study SN spectral variations and evolution. We expect that
this will help substantially in calibrating our use of SNe as distance
indicators and will allow improved calibration of the SDSS-II and future
larger (e.g., the Panoramic Survey Telescope and Rapid Response System, the
Dark Energy Survey, the Large Survey Telescope, and the Joint Dark Energy
Mission) SN samples.

\section*{ACKNOWLEDGMENTS}

Funding for the SDSS and SDSS-II has been provided by the Alfred P. Sloan
Foundation, the Participating Institutions, the National Science Foundation 
(NSF), the U.S. Department of Energy, the National Aeronautics and Space
Administration (NASA), the Japanese Monbukagakusho, the Max Planck Society, 
and the Higher Education Funding Council for England. The SDSS Web Site is
http://www.sdss.org/.

The SDSS is managed by the Astrophysical Research Consortium for the
Participating Institutions. The Participating Institutions are the American
Museum of Natural History, Astrophysical Institute Potsdam, University of
Basel, University of Cambridge, Case Western Reserve University, University of
Chicago, Drexel University, Fermilab, the Institute for Advanced Study, the
Japan Participation Group, Johns Hopkins University, the Joint Institute for
Nuclear Astrophysics, the Kavli Institute for Particle Astrophysics and
Cosmology, the Korean Scientist Group, the Chinese Academy of Sciences
(LAMOST), Los Alamos National Laboratory, the Max-Planck-Institute for
Astronomy (MPIA), the Max-Planck-Institute for Astrophysics (MPA), New Mexico
State University, Ohio State University, University of Pittsburgh, University
of Portsmouth, Princeton University, the United States Naval Observatory, and
the University of Washington.

This work is based in part on observations made at the following
telescopes. The Hobby-Eberly Telescope (HET) is a joint project of the
University of Texas at Austin, the Pennsylvania State University, Stanford
University, Ludwig-Maximillians-Universit\"{a}t M\"{u}nchen, and 
Georg-August-Universit\"{a}t G\"{o}ttingen. The HET is named in honor of its 
principal benefactors, William P. Hobby and Robert E. Eberly. The Marcario
Low-Resolution Spectrograph is named for Mike Marcario of High Lonesome
Optics, who fabricated several optical elements for the instrument but died
before its completion; it is a joint project of the Hobby-Eberly Telescope
partnership and the Instituto de AstronomǛa de la Universidad Nacional
Autonoma de Mexico. We thank the HET resident astronomers (John Caldwell,
Heinz Edelmann, Steve Odewahn, and Matthew Shetrone) for their continued
effort and support with the HET observations. The Apache Point Observatory
3.5-m telescope is owned and operated by the Astrophysical Research
Consortium. We thank the observatory director, Suzanne Hawley, and site
manager, Bruce Gillespie, for their support of this project. The Subaru
Telescope is operated by the National Astronomical Observatory of Japan. The
William Herschel Telescope is operated by the Isaac Newton Group, and the
Nordic Optical Telescope is operated jointly by Denmark, Finland, Iceland,
Norway, and Sweden, both on the island of La Palma in the Spanish Observatorio
del Roque de los Muchachos of the Instituto de Astrofisica de
Canarias. Observations at the ESO New Technology Telescope at La Silla
Observatory were made under programme IDs 77.A-0437, 78.A-0325, and 
79.A-0715. Kitt Peak National Observatory, National Optical Astronomy Observatories
(NOAO), is operated by the Association of Universities for Research in Astronomy,
Inc. (AURA) under cooperative agreement with the NSF. 
The WIYN Observatory is a joint facility of the University of
Wisconsin-Madison, Indiana University, Yale University, and NOAO.
The W. M. Keck Observatory is operated as a
scientific partnership among the California Institute of Technology, the
University of California, and NASA.
The Observatory was made possible by the generous financial
support of the W. M. Keck Foundation. The South African Large Telescope of the
South African Astronomical Observatory is operated by a partnership between
the National Research Foundation of South Africa, Nicolaus Copernicus
Astronomical Center of the Polish Academy of Sciences, the Hobby-Eberly
Telescope Board, Rutgers University, Georg-August-Universit\"{a}t G\"{o}ttingen,
University of Wisconsin-Madison, University of Canterbury, University of North
Carolina-Chapel Hill, Dartmouth College, Carnegie Mellon University, and the
United Kingdom SALT consortium. A.V.F.'s supernova group at U.C. Berkeley is
supported by NSF grant AST--0607485.

This work is also supported in part by the U.S. Department of Energy under contract number DE-AC0276SF00515.

\begin{deluxetable}{cccrcr}
\tabletypesize{\scriptsize}
\tablecaption{SDSS-II SN Spectroscopic Follow-up Observations\label{tbl:obs}}
\tablewidth{0pt}
\tablehead{
\colhead{IAU \tablenotemark{a}} & \colhead{} & \colhead{R.A.} & \colhead{Decl.} & \colhead{} & \colhead{Obs Date} \\
\colhead{Name} & \colhead{SNID\tablenotemark{b}} & \colhead{(J2000)} & \colhead{(J2000)} & \colhead{Telescope} & \colhead{(UT)} 
}
\startdata
2005ed	&	722	&	00	02	49.36	&	+	00	45	04.8	&	APO	&	9/9/05	\\
2005ef	&	739	&	00	58	22.87	&	+	00	40	44.6	&	APO	&	9/9/05	\\
2005ei	&	744	&	21	56	47.65	&	+	00	19	03.0	&	APO	&	9/8/05	\\
2005eg	&	762	&	01	02	08.49	&	-	00	52	44.5	&	APO	&	9/9/05	\\
2005ex	&	774	&	01	41	51.24	&	-	00	52	35.0	&	APO	&	9/9/05	\\
2005ex	&	774	&	01	41	51.24	&	-	00	52	35.0	&	MDM	&	9/15/05	\\
---	&	779	&	01	46	41.70	&	-	01	01	14.1	&	HET	&	12/29/05	\\
2005ez	&	1032	&	03	07	10.97	&	+	01	07	10.4	&	APO	&	9/25/05	\\
\enddata

\tablecomments{}
\tablenotetext{a}{Official IAU name, when assigned.}
\tablenotetext{b}{SDSS Internal SN candidate ID.}
\tablenotetext{c}{Spectroscopically confirmed by another group using the HET \citep[CBET 266;][]{qu07}. The spectrum is not available.}
\tablenotetext{d}{Spectroscopically confirmed by the ESSENCE group.}
\end{deluxetable}

\clearpage

\begin{deluxetable}{ccrrcrcrrcr}
\tabletypesize{\scriptsize}
\tablecaption{SDSS-II SN Redshift Cross-Correlation and Type Analysis \label{tbl:spec}}
\tablewidth{0pt}
\tablehead{
\colhead{IAUC ID \tablenotemark{a}} & \colhead{SNID\tablenotemark{b}} & \colhead{Epoch$_{\rm p}$\tablenotemark{c}} & \colhead{Epoch$_{\rm s}$\tablenotemark{d}} & \colhead{Type$_{\rm SN}$\tablenotemark{e}} & \colhead{$z_{SN}$\tablenotemark{f}} & \colhead{SN Temp\tablenotemark{g,h}} & \colhead{$r_{SN}$\tablenotemark{i}} & \colhead{$z_g$\tablenotemark{j}} & \colhead{Type$_{\rm g}$\tablenotemark{k}} & \colhead{$r_g$\tablenotemark{l}}
}
\startdata
2005ed 	& 	722	& 	14	& 	17	& 	Ia 	& 	0.087	& 	 	& 	11.9	& 	0.0861	& 	E/S0 	& 	17.4	\\
2005ef 	& 	739	& 	15	& 	16	& 	Ia 	& 	0.105	& 	 	& 	12.9	& 	0.107	& 	E/S0 	& 	16.5	\\
2005ei 	& 	744	& 	11	& 	8	& 	Ia 	& 	0.123	& 	 	& 	4.4	& 	0.1282	& 	Sm/Im 	& 	4.7	\\
2005eg 	& 	762	& 	0	& 	-1	& 	Ia 	& 	0.189	& 	 	& 	7.5	& 	0.1908	& 	Sa 	& 	13.3	\\
2005ex 	& 	774	& 	14	& 	16	& 	Ia 	& 	0.09	& 	 	& 	10.4	& 	0.0933	& 	Sa 	& 	9.5	\\
2005ex 	& 	774	& 	22	& 	 	& 	Galaxy 	& 	 	& 	 	& 	 	& 	0.0936	& 	Sa 	& 	12	\\
--- 	& 	779	& 	110	& 	 	& 	Galaxy 	& 	 	& 	 	& 	 	& 	0.2377	& 	Sbc/Sc 	& 	16.8	\\
2005ez 	& 	1032	& 	14	& 	22	& 	Ia 	& 	0.133	& 	 	& 	8.4	& 	0.1293	& 	Sa 	& 	10.3	\\
\enddata
\end{deluxetable}

\clearpage

\begin{deluxetable}{ccc||c||ccccc}
\tabletypesize{\scriptsize}
\tablecaption{SDSS-II SNe Spectral Decomposition and Host Analysis Results \label{tbl:host}}
\tablewidth{0pt}
\tablehead{
\colhead{SNID\tablenotemark{a}} & \colhead{Tel\tablenotemark{b}} & \colhead{Epoch$_p$\tablenotemark{c}} & \colhead{SN$_s$\tablenotemark{d}} & \colhead{Epoch$_{sh}$\tablenotemark{e}} & \colhead{Host$_{sh}$\tablenotemark{f}} &\colhead{SN Frac\tablenotemark{g}} & \colhead{$\chi^2_{\nu}$\tablenotemark{h}} & \colhead{$Q$\tablenotemark{i}}
}
\startdata
1112	&	HET	&	10	&	Ia	/	1999ac	&	9	&	Sb	&	0.307	$\pm$	0.021	&	1.88	&	-1	\\
1253	&	HET	&	5	&	Ia	/	1999ac	&	7	&	Sb	&	0.669	$\pm$	0.019	&	2.14	&	2.6	\\
2017	&	HET	&	3	&	Ia	/	1991T	&	-5	&	Sbc/Sc	&	0.351	$\pm$	0.018	&	1.96	&	-0.5	\\
2030	&	HET	&	29	&	Ia	/	1990I	&	44	&	Sbc/Sc	&	0.431	$\pm$	0.059	&	4.93	&	-0.1	\\
2533	&	HET	&	8	&	Ia	/	1991T	&	2	&	E/S0	&	0.584	$\pm$	0.03	&	0.88	&	4.4	\\
3199	&	HET	&	8	&	Ia	/	1997cy	&	11	&	Sbc/Sc	&	0.499	$\pm$	0.084	&	2.12	&	-0.6	\\
3377	&	HET	&	13	&	Ia	/	1999ac	&	8	&	Sm/Im	&	0.743	$\pm$	0.032	&	2.54	&	-0.2	\\
4000	&	HET	&	17	&	Ia	/	1997br	&	14	&	Sbc/Sc	&	0.395	$\pm$	0.033	&	1.39	&	0	\\
\enddata
\tablecomments{}
\tablenotetext{a}{SDSS internal SN ID.}
\tablenotetext{b}{Spectroscopic follow-up telescope.}
\tablenotetext{c}{Observation date of the SN relative to $B$-band maximum brightness estimated from light-curve fits.}
\tablenotetext{d}{The best and second-best SN type from cross-correlation analysis. 
``Ia?'' indicates best fit to a SN Ia template, but a reasonable fit to the next-best type.}
\tablenotetext{e}{Age of the SN relative to $B$-band maximum brightness estimated from spectrum decomposition.}
\tablenotetext{f}{Host-galaxy type ID from spectral decomposition. Unknown indicates too little host light 
for type ID. }
\tablenotetext{g}{Estimate of the SN light fraction in the observed spectrum, with statistical error (the maximum value of the asymmetric errors obtained from the PCA code).}
\tablenotetext{h}{Reduced $\chi^2$ value.}
\tablenotetext{i}{Quality factor $Q=\frac{\Delta \chi^2_{\nu}\sqrt{\nu}}{\chi^2_{\nu}}$, 
which estimates the significance of the $\chi^2$ differentiation between the best two supernova
types. Negative values indicate a selected type differing from that determined by the
cross-correlation analysis. }
\tablenotetext{j}{SNe above the line have extinction estimates from MCLS2k2 analysis \citep{jh07}.
Those below the line have $A_V$ from light-curve fitting using the frame difference photometry.}

\end{deluxetable}
\end{document}